\documentclass[aps,prd,preprintnumbers, nofootinbib, onecolumn]{revtex4}
\usepackage{bm}
\usepackage{latexsym}
\usepackage{amsmath,amsfonts,amssymb}
\usepackage{graphicx,epsfig}
\usepackage{psfrag}
\usepackage{amsthm}
\usepackage{bigints}
\usepackage{xcolor}
\usepackage{amsmath}
\usepackage{braket}
\usepackage{indent first}
\usepackage[normalem]{ulem}
\usepackage{fancyhdr}
\usepackage{multirow}
\useunder{\uline}{\ul}{}
 \usepackage{environ}
\usepackage{mathtools}
\usepackage{float}
\usepackage{relsize}

\def\o{\omega}

\def\a{\alpha}
\def\b{\beta}

\def\d{\delta}
\def\D{\Delta}

\def\t{\tau}
 
\def\p{\partial}
\def\f{\frac}

\newcommand{\be}{\begin{equation}}
\newcommand{\ee}{\end{equation}}
\newcommand{\bes}{\begin{equation*}}
\newcommand{\ees}{\end{equation*}}

\newcommand{\beq}{\begin{eqnarray}}
\newcommand{\eeq}{\end{eqnarray}}

\NewEnviron{eqn}{
\begin{align}
\begin{split}
  \BODY
\end{split}
\end{align}
}

\NewEnviron{eqn*}{
\begin{align*}
\begin{split}
  \BODY
\end{split}
\end{align*}
}

\newcommand{\tom}{\tilde{\o}}
\begin{document}
\title{Unruh-Fulling effect in nonlocal field theory: The role of Unruh decomposition}

\author{Ashmita Das}
\email{ashmita.phy@gmail.com}
\affiliation{School of Physical Sciences, Indian Association for the Cultivation of Science, Kolkata 700032, India \& Department of Theoretical Sciences, S N Bose National Centre for basic Sciences, Kolkata 700106, India}

\author{Bibhas Ranjan Majhi}
\email{bibhas.majhi@iitg.ac.in }
\affiliation{Department of Physics, Indian Institute of Technology Guwahati, Guwahati 781039, Assam, India}
%%%
%%%%%%%%%%%%%%%%%
\begin{abstract}
%%%%%%%%%%%%%%%%%
%
%
%
%
\par\noindent
We investigate the Unruh-Fulling effect in a class of nonlocal field theories by examining both the number operator and Unruh-DeWitt detector methods. Unlike in previous literature, we use Unruh quantization to quantize the matter field. Such choice, as oppose to standard Minkowski decomposition, naturally incorporates the time translational invariance in the positive frequency Wightman function and thus captures the thermal equilibrium of the system. We analyze the Unruh-Fulling effect for a massless real scalar field in both the Lorentz noninvariant and Lorentz invariant nonlocal theories. In Lorentz noninvariant nonlocal theory, the expectation value of number operator and the response function of the detector are modified by an overall multiplicative factor. Whereas in Lorentz invariant nonlocal theory these quantities remain identical to those of the standard Unruh-Fulling effect. The temperature of the thermal bath remains unaltered for both the Lorentz noninvariant and Lorentz invariant nonlocal theories. Therefore, in terms of temperature, the nonlocal Unruh-Fulling effect is universal while it is derived via Unruh quantization, whereas the transition rate may be modified.
\end{abstract}
\maketitle
\section{Introduction}\label{introduction}
Observer dependent results of fundamental physics have always fascinated us in their own right. Among many such theories, the idea of the Unruh-Fulling (UF) effect has unveiled one of the interesting features of quantum field theory (QFT). It says that in Minkowski spacetime a uniformly accelerating observer (called Rindler observer) perceives the vacuum state of a field as a thermal bath in equilibrium with temperature $T=a/2\pi$, where a symbolizes the acceleration of the observer  \cite{Fulling:1972md,DeWitt:1975ys,Unruh:1976db}. 
The UF effect is an important tool to explore the Hawking radiation in black hole spacetime \cite{Hawking:1974rv,Hawking:1975vcx}, particle emission by cosmological horizon \cite{Gibbons:1977mu}, and particle creation in curved spacetime  \cite{Hajicek}. The phenomenology of the UF effect can be explored by a uniformly accelerated two level atomic detector (UD detector) and its transition rate/power spectrum \cite{Hawking:1979ig,Grove:1983rp}.
We have ample literature in the context of UD detector and its implications in the gravitational physics, cosmology, condensed matter physics, quantum optics, quantum entanglement, entanglement harvesting (see \cite{Oriti:1999xq,Lenz:2008vw,Sudhir:2021lpf,Toussaint:2021czo,Gooding:2020scc,Scully:2003zz,Scully:2017utk,Fuentes-Schuller:2004iaz,Reznik:2002fz,Lin:2010zzb,1,2,3,4} and the references therein).
In the experimental ventures of the UF effect,  the temperature of the thermal bath plays the key role. 
%%%
Any deviation in the temperature of the bath from the standard Unruh temperature
could possibly be the signature of modified  QFT. 
%%%%
In this regard, there is a long-standing debate about whether the Unruh temperature gets any modification in a nonlocal field theory.
Nonlocal field theories were initially introduced to remove the ultraviolet divergences as they appear in a local theory of quantised fields \cite{Pais:1950za}.
Nonlocality may emerge as a low energy phenomena in the effective interaction. It can also be originated from the fundamental theories such as loop quantum gravity \cite{Rovelli:1989za,Rovelli:1991zi} and string theory \cite{Mende:1992pm,Strominger:1991iy}.  
Nonlocal theories have significant impacts in gravitational physics, cosmology, quantum theory of gravity (QG), standard model of particle physics and QFT (see \cite{Modesto:2011kw,Modesto:2014lga,Douglas:2001ba} for detailed discussion). 
There are two popular versions of nonlocal field theories. In one kind the nonlocality is represented by a length scale which is not required to be a minimal length [Lorentz invariant nonlocal theory (LI- NLT)]. Whereas, the other version contains a minimal length scale [Lorentz noninvariant nonlocal theory (LNI- NLT)].
LI-NLT has advantage due to its Lorentz invariant character;  however it eludes us from the origin of such arbitrary length scale in the theory \cite{Douglas:2001ba,Modesto:2011kw,Modesto:2014lga}.
 LNI-NLT destroys the Lorentz invariance, although it intuitively develops how the inclusion of QG effects lead to a minimal length scale of the order of the Planck length $(l_{\rm Pl})$ \cite{Amelino-Camelia:2000cpa,Amelino-Camelia:2000stu}.
The LNI-NLTs must be accommodated within a modified version of Einstein's special relativity, which is popularly known as doubly special relativity (DSR) \cite{Magueijo:2001cr,Magueijo:2002am}. It postulates that along with the speed of light the minimal length scale (i.e $l_{\rm Pl}$) also remain as an observer (inertial) independent quantity. 
Several thought experiments and fundamental theories of QG support the existence of a minimal length scale 
%($\mathcal{O}(l_{\rm Pl})$) 
(please see \cite{Garay:1994en} and the references therein). 

{\it What are the debates regarding the UF effect in nonlocal theories?} 
%In 2011, with the work of Nicoloni {\it et. al.} \cite{Nicolini:2009dr}, a series of debates has begun in this context. 
In \cite{Nicolini:2009dr}, Nicoloni {\it et. al.} have shown that in a nonlocal theory with a minimal length scale, the transition rate of the UD detector and the Unruh bath temperature get significantly enhanced.
The authors have considered a Lorentz invariant length scale ($l$) in the theory without assuming any particular value for it \cite{Nicolini:2009dr}. $l$ may be determined from a fundamental theory of QG where in all probability it can be chosen as $l_{Pl} \sim 10^{-35}$ m \cite{Nicolini:2009dr}. 
Later, it was revealed in \cite{Kajuri:2017jmy} that for a wide class of nonlocal theories, the UF effect remains unchanged 
when derived via the Bogoliubov transformation method (the same conclusion has drawn for Hawking effect as well \cite{Kajuri:2018myh}). This is in contradiction to the outcome of \cite{Nicolini:2009dr}.
The author also argued that in a nonlocal theory the Bogoliubov coefficient method and UD detector method produce inequivalent results as the detector method assumes a point like interaction despite the nonlocality of the theory \cite{Kajuri:2017jmy}.
 Thus two following conflicting aspects arose: i) in a nonlocal theory, the UF effect remains unchanged or not? ii) If it remains intact why do the two standard methods which otherwise produce similar outcomes yield two different results? 
In \cite{Modesto:2017ycz}, the authors 
have agreed upon the consistency of nonlocal UF effect as reported in \cite{Kajuri:2017jmy}. They have also pointed out a 
misapprehension in defining the Wightman function in \cite{Nicolini:2009dr}, which 
caused the ``misleading enhancement" in the transition rate of the detector.
In addition, the authors have opposed the argument for the inequivalent outcome of the two methods \cite{Kajuri:2017jmy} and shown  that for a general class of weekly nonlocal theory, the UD detector method produces the same result as the Bogoliubov coefficient one \cite{Modesto:2017ycz}. 
Note that in \cite{Kajuri:2017jmy,Modesto:2017ycz}, a class of LI-NLT have been considered. 

In succession to \cite{Modesto:2017ycz}, W. Kim {\it et. al.} have studied the 
UF effect in a class of LNI-NLT, 
where their results suggest that the UF effect and the bath temperature acquire modifications due to the nonlocality \cite {Gim:2018rcy}. 
The response function 
of the detector explicitly depends on the proper time of the detector which ruins the global thermal equilibrium of the Minkowski vacuum as perceived by the Rindler observer
\cite {Gim:2018rcy}.
In other words the nonlocal Wightman function gives up its time translational symmetry with respect to the Rindler proper time. Therefore this nonlocal Wightman function does not satisfy the Kubo-Martin-Schwinger (KMS) condition \cite{Kubo:1957mj,Martin:1959jp} which guarantees that the detector and the quantum field are not in thermal equilibrium.
This finding has triggered another debate about whether the consistency of the UF effect depends on the nature of the nonlocal theories.

{\it Our motivation and further proceedings:} We enlist the accepted results in the context of nonlocal UF effect \cite{Kajuri:2017jmy,Modesto:2017ycz,Gim:2018rcy} as follows.
\begin{itemize}
\item The LI-NLT preserves the time translational invariance of the Wightman function (and so preserves the thermal equilibrium of the system) in terms of the Rindler proper time. This has been calculated by using the usual Minkowski mode decomposition in \cite{Modesto:2017ycz}. Furthermore the equivalence between the two standard methods of studying the UF effect was established \cite{Modesto:2017ycz}.
\item LNI-NLT case is discussed in \cite{Gim:2018rcy} by following only the detector's response method. The Wightman function as derived via Minkowski quantization, turns out to be time translationally noninvariant along the Rindler trajectory. This destroys the thermal equilibrium of the system.  
\end{itemize}
There is evidence \cite{Barman:2021oum,Chowdhury:2019set} that even in the local field theory, under certain conditions (e.g. thermal field Wightman function in a uniformly accelerated frame or in a uniformly rotating frame), the Minkowski quantization yields non-equilibrium condition while breaking the time translational invariance in Wightman function.

A Thermodynamic definition of temperature is well defined in equilibrium condition. Although there have been a few attempts (see \cite{Jou:2003}) to define an effective temperature in a nonequilibrium system, they lack a concrete theoretical framework.
Therefore the identification of thermodynamic quantities as obtained in \cite{Gim:2018rcy} are not free of ambiguity.
Unruh showed in \cite{Unruh:1976db} that a particular combination of right and left Rindler modes gives rise to a well-behaved,  {\it i.e.} analytic and bounded, solution to the field. 
These are known as Unruh modes in literature. Unruh modes share the same vacuum as the Minkowski modes and thus the  Unruh annihilation operator annihilates the Minkowski states \cite{Unruh:1976db,birrell}. 
Similarly, the Rindler creation and annihilation operators can be written in terms of the Unruh creation and annihilation operators \cite{Unruh:1976db,birrell}.
In various instances (e.g. thermal Wightman function in Rindler frame \cite{Barman:2021oum,1,Chowdhury:2021ieg}), it has been observed that the Wightman functions have a natural time translational invariance with respect to the Rindler proper time when derived using the Unruh mode decomposition. 
This makes the Unruh quantization a useful tool to maintain the thermal equilibrium in accelerated frame. This thermal equilibrium aids in determining the thermodynamical entities (temperature, entropy etc.) without any ambiguity. 
%Hence the use of Unruh quantization becomes a natural choice in the relevant circumstances. 
%
%
As in LNI-NLT, the Minkowski quantization destroys the thermal equillibrium \cite{Gim:2018rcy}, therefore
we expect that the same can be restored by implementing the Unruh quantization. 

Therefore in the present manuscript we aim to explore how the choice of the mode solutions and quantization rules govern the UF effect? 
The same classes of LNI-NLT as in \cite{Gim:2018rcy} and LI-NLT in \cite{Kajuri:2017jmy,Modesto:2017ycz} have been considered in our work. We examine the nonlocal UF effect via the UD detector and number operator methods. Unlike \cite{Gim:2018rcy, Kajuri:2017jmy,Modesto:2017ycz}, we use Unruh quantization and show that the nonlocal Wightman function becomes time translationally invariant  with respect to the Rindler proper time. Therefore the required thermal equilibrium is achieved in LNI-NLT.  We also briefly discuss LI-NLT in this context. 
To determine the 
temperature of the thermal bath we show that the nonlocal Wightman function and the ratio of the excitation and deexcitation probabilities of the detector satisfy the detailed balance form of the KMS condition (see \cite{Fewster:2016ewy,Garay:2016cpf} and the relevant references therein).
This implies that irrespective of the LI-NLT or LNI-NLT the temperature of the thermal bath stands unaltered from that of the standard Unruh temperature ($a/2\pi$).
These outcomes are in sharp contrast from the results in \cite {Gim:2018rcy}. Although the temperature of the thermal bath remains unchanged, the nonlocal Wightman function, transition rate of the detector and the expectation value of the number operator, acquire a modification by the form factor of the LNI-NLT.  For LI-NLT the results are exactly matched with \cite{Modesto:2017ycz}. 
We feel that our work has brought in another angle to this ongoing debate: how does the UF effect depend on the choice of mode solutions and quantization rules in a nonlocal theory?

We present our work in the following order. In Sec. \ref{model}, we describe the nonlocal theory as considered in the present manuscript. We obtain the nonlocal Unruh mode solutions and Wightman function in the accelerated frame in Sec.  \ref{numberop_unmode1}. The next section contains the nonlocal UF effect via the two standard methods. Section \ref{detailed balance_1} depicts the analysis of thermality and temperature of thermal bath. A discussion related to LI-NLT is presented in Sec.  \ref{LI}. In Sec.  \ref{discussions} we discuss our findings. Appendices are provided at the end to be self-sufficient. 
%%%
\subsection{Summary of symbols}
%%%
Before proceeding further, we introduce the relevant local and nonlocal parameters as follows,
\begin{center}
\begin{tabular}{ |c|c|c|c|c| } 
\hline
Parameters / operators & Local version  & nonlocal version & vacuum states\\
\hline
Minkowski annihilation operator & $a_k$ & $g_k$ & $\ket{0}_M$  \\ 
\hline
Rindler annihilation operator & $b_k$ & $c_k$ & $\ket{0}_R$\\ 
\hline
Unruh annihilation operator & $d_k$ & $p_k$ & $\ket{0}_M$\\
\hline
\end{tabular}
\end{center}
%%%

%% 
\section{model description: LNI-NLT}\label{model}
%%%%
We consider a class of nonlocal theories which involve a minimal length scale ($l$) and higher order derivative terms.
These theories violate Lorentz invariance ({\it i.e} LNI-NLT) and can be accommodated within the DSR.  In this background we write the action for a nonlocal massless scalar field as follows \cite{Kimberly:2003hp, Gim:2018rcy}:
%%%
\beq
S_{{\rm NL}}=\,\int\,d^4x\, \sqrt{-g}\,\bigg(-\f{1}{2}\bigg)\,\phi_{{\rm NL}}(x)\,[-\Box\,f\,((i\,l\,\p_{0})^2,\,(i\,l\,\p_{i})^2)]\,\phi_{{\rm NL}},
\label{action_nl_1}
\eeq
%%%
where $f$ is an analytic and nonzero function at every point of the spacetime and $\phi_{{\rm NL}}$ is a massless scalar field in the nonlocal theory. 
From the above action the field equation becomes, 
%%%%
\beq
\Box\,f\,((i\,l\,\p_{0})^2,\,(i\,l\,\p_{i})^2)\,\,\phi_{{\rm NL}}=\,0~.
\label{scalar_nonl_1}
\eeq
The solution for $\phi_{\textrm{NL}}$ can be obtained from those of local field $\phi$ as below,
\beq
\phi_{\rm NL}=\,f^{-1}\,((i\,l\,\p_{0})^2,\,(i\,l\,\p_{i})^2)\,\phi~,
\label{scalar_loc_nl}
\eeq
which satisfies Eq. (\ref{scalar_nonl_1}) and the local field $\phi$ obeys $\Box\phi=0$~.
 The appearance of the ghost fields in these kinds of nonlocal theories has been tackled by examining the pole structure \cite{Gim:2018rcy}. It can be seen that 
as long as $f$ is analytic and nonzero at every point of spacetime, the pole structure is not altered from that of the local massless scalar field theory. This ensures the nonappearance of the ghostlike excitations in the present nonlocal model. 
 The canonical momentum as defined in the local field theory suggests that the same can be written for the LNI-NLT as  $\Pi_{\rm NL}=\,\f{\d S_{\rm NL}}{\d \dot{\phi}_{\rm NL}}$. Nevertheless the nonlocality of the model restricts one to write the  $\Pi_{\rm NL}$ solely in terms of $\dot{\phi}_{\rm NL}$ \cite{Gim:2018rcy}. Thus the Hamiltonian corresponding to the LNI-NLT cannot be obtained from the Lagrangian by following the standard approaches. In addition the stabilization issue of the quantum description of LNI-NLT is a subject of in depth examination and should be addressed by following the prescriptions as developed in \cite{Gomis:2000gy,Gomis:2003xv,Barnaby:2007ve}. Nonetheless we will follow here the existing progress of this theory in studying the UF effect (see \cite{Kajuri:2017jmy,Kajuri:2018myh,Modesto:2017ycz,Gim:2018rcy}) where the field Hamiltonian has no such role.

It is discussed in literature \cite{Barnaby:2010kx,Gim:2018rcy} that the number of independent solutions for an infinite order differential equation [e.g. Eq. (\ref{scalar_nonl_1})] depend on the number of poles that appear in its propagator. 
Currently discussed nonlocal models possess the same number of poles in the propagator 
as the local quantum field equation. Therefore the nonlocal 
field solutions turn out to be the same as  corresponding to $\Box\,\phi=\,0$. 
Therefore the plane wave solutions viz. $\sim e^{\pm ik_{\mu}x^\mu}$ can be considered as the solutions for the nonlocal field equation (Eq. (\ref{scalar_nonl_1})). 
The notation $k_\mu$ in the Minkowski spacetime is $k_{\mu}=(\o,\,k_i)$ where $i=1,\,2,\,3$ in $(1+3)$ dimensions.
Note that the Lorentz violating effects of these nonlocal theories make an appearance through the creation and the annihilation operators which leads to the modifications in the operator sector instead affecting the mode solutions. As a consequence the commutation relations between the operators get altered from the local one  \cite{Gim:2018rcy}. 
 As the mode solutions of $\phi_{\rm NL}(x)$ do not get any modifications due to the nonlocality, thus plugging the plane wave solutions in Eq.(\ref{scalar_nonl_1}), one obtains the dispersion relation as, $f(l^2\,\o^{2}, l^2\,k_{i}^{2})\,k_{\mu}k^{\mu}=\,0$.
Below we give a brief discussion on the detector's response function which is well reviewed in literature \cite{birrell, Crispino}. We consider a uniformly accelerated two level atomic detector in the right Rindler wedge (RRW) interacting with a massless scalar field in the background of a LNI-NLT, where the detector is represented by a monopole. 
We also take the adiabatic switching function for the detector to be of the form $\sim e^{s|\t|}$, where $\t$ represents the proper time of the detector. We eventually take $s \to 0$ which leads to an infinite interaction time between the field and the detector. Using the Unruh quantization the response function per unit time of the detector comes out to be as follows  \cite{birrell, Crispino}:
\beq
I^{R}(\D E)=\,\int_{-\infty}^{\infty}\,d (\D \t)\,e^{-i\D E \D \t}\,\,{}^{R}G_{W}(\D \t)~.
\label{power_1}
\eeq
Here, $\Delta E$ and $\Delta\tau$ denote the energy difference between two states of the detector and difference in proper time ($\tau$) of the same, respectively. ${}^{R}G_{W}(\D \t)$ is the positive frequency Wightman function in the RRW.
In Eq. (\ref{power_1}), the Wightman function depends on the duration of the interaction $(\Delta \t)$ but not the explicit values of the initial and final times. Therefore the rate of transition probability remains constant for a fixed time interval irrespective of the value of the initial or final time $\tau_0$. This leads to the condition of thermal equilibrium between the detector and the field.  In \cite{Gim:2018rcy} the nonlocal Wightman function depends explicitly on initial and final times and thus becomes time translationally noninvariant. 
Therefore the above definition of transition rate does not provide the equilibrium situation and puts ambiguity in the definition of temperature.
We suspect that whenever the Wightman function depends explicitly on time, defining the transition rate as in Eq. (\ref{power_1}) is not a robust one . 
In the upcoming section we derive the nonlocal Unruh mode solutions of the scalar field in $(1+3)$ dimensions.
For $(1+1)$ dimensional case we refer to Appendices \ref{mod_rindler}, \ref{numberop_unmode} and \ref{response_2d_nonlocal}. 
\section{Nonlocal field in terms of the Unruh operators and Unruh modes}\label{numberop_unmode1}
%%%%%
%
The local Rindler mode solutions as depicted in the Eqs. (\ref{nonl_rrw_2}) and (\ref{nonl_lrw_2}), are individually nonanalytic in the whole Minkowski spacetime. Whereas, the Unruh modes which are formed by combining both the Rindler modes such as ${}^{R}u_{k}+\,e^{-\f{\pi \tom}{a}}\,{}^{L}u_{-k}^{*}$ and ${}^{R}u^{*}_{-k}+\,e^{\f{\pi \tom}{a}}\,{}^{L}u_{k}$
are analytic and bounded in the whole Minkowski spacetime \cite{Unruh:1976db,birrell}. 
Our readers may follow \cite{birrell} for a discussion on the formation of Unruh mode solutions and their properties. 
In ($1+1$) and  ($1+3$) dimensional Rindler quantization the $(n-1)$ tuple coordinates are  $\tom=|k|$ and $(\tom, k_{x}, k_{y})=\,(\tom,k_{\bot})$ respectively, where $n$ represents the dimensions of spacetime. 
In $(1+3)$ dimensions the Unruh mode solutions for a nonlocal field equation can be found by using Eq. (\ref{scalar_loc_nl}). 
%For the $(1+1)$ dimensional case we refer our readers to the Appendix \ref{numberop_unmode}. 
%
%
The field solutions corresponding to $\Box \phi=\,0$ can be written in terms of the Unruh modes (so in terms of Rindler mode) and Unruh operators as follows \cite{Crispino}: 
\beq
\phi(x)&=&\,\phi^{R}+\,\phi^L\nonumber\\
&=&\,\sum_{\tom=0}^{\infty}\sum_{k_{\bot}=-\infty}^{\infty}\f{1}{\sqrt{2\,{\rm Sinh}\,(\f{\pi \tom}{a})}}\,\bigg[\bigg\{d^{1}_{k}\,e^{\f{\pi\,\tom}{2\,a}}\,{}^{R}u_{\tom k_{\bot}}(\eta,\xi,x_{\bot})+\,d^{2}_{k}\,e^{-\f{\pi\,\tom}{2\,a}}\,^{R}u^{*}_{\tom -k_{\bot}}(\eta,\xi,x_{\bot})\bigg\}\nonumber\\
&+&\,\bigg\{d^{1}_{k}\,e^{-\f{\pi\,\tom}{2\,a}}\,^{L}u^{*}_{\tom k_{\bot}}(\eta,\xi,x_{\bot}))
+\,d^{2}_{k}\,e^{\f{\pi\,\tom}{2\,a}}\,^{L}u_{\tom k_{\bot}}(\eta,\xi,x_{\bot}))\bigg\}\bigg]+\,H.c.~.
\label{unruh_op_3d_1}
\eeq
In the above $H.c.$ refers to hermitian conjugate.
$d^{(1,\,2)}$ are known as the Unruh operators in QFT. The Rindler modes for a massless scalar field in local theory can be written as \cite{Crispino}, 
%%%
\beq
{}^{R,L}u_{\tom k_{\bot}}(\eta,\xi,x_{\bot})=\,\bigg[\f{{\rm Sinh}(\pi \tom/a)}{4 \pi^4\,a}\bigg]^{1/2}\,K_{\f{i\tom}{a}}\bigg(\f{|k_\bot |\,e^{a\xi}}{a}\bigg)\,e^{i k_\bot.x_{\bot}\,\mp\,i\tom \eta}~.
\label{rindler_3d_1}
\eeq
%%%
The Unruh mode solutions possess the positive frequency analyticity properties corresponding to Minkowski time, which ensures that the Unruh annihilation operators also annihilate the Minkowski vacuum state such as $d^{1}_{k}\ket{0_M}=d^{2}_{k}\ket{0_M}=\,0$.
 In general, the purpose of using the Unruh quantization in studying the UF effect, can be realized in the following ways. 
 The Unruh operators act in the same manner as the Minkowski operators. 
 Also the Unruh mode solutions are the specific combination of Rindler modes which correspond to the accelerated observer. Therefore the construction of Unruh quantization supports the feature of an accelerated observer and an inertial (Minkowski) vacuum state as well. 
%${}^{R}u_{k}$ and ${}^{L}u_{k}$ represent the wave solution in the right and left Rindler Wedge 
The Unruh operators are related to the Rindler creation (${}^{R,L}b_{k}^{\dagger}$) and annihilation operators $({}^{R,L}b_k)$ as below:  
%%%%%%
\beq
{}^{L}b_{k}=\,\f{1}{\sqrt{2\,{\rm Sinh}\,(\f{\pi \tom}{a})}}\,\bigg[e^{\f{\pi\,\tom}{2\,a}}\,d_{k}^{2}+\,e^{-\f{\pi\,\tom}{2\,a}}\,d^{1^{\dagger}}_{-k}\bigg]~;
\label{BRM2}
\\
{}^{R}b_{k}=\,\f{1}{\sqrt{2\,{\rm Sinh}\,(\f{\pi \tom}{a})}}\,\bigg[e^{\f{\pi\,\tom}{2\,a}}\,d_{k}^{1}+\,e^{-\f{\pi\,\tom}{2\,a}}\,d^{2^{\dagger}}_{-k}\bigg]~.
\label{rindler_unruh_rrw}
\eeq
It is worth recalling that Rindler annihilation operator annihilates the Rindler vacuum state $(\ket{0}_R)$, such as ${}^{R,L}b_k \ket{0}_R=\,0$.
Using Eq. (\ref{scalar_loc_nl}) we find the nonlocal field solutions in terms of the Unruh modes as,
\beq
\phi_{\rm NL}=\, \phi^R_{\textrm{NL}} + \phi^L_{\textrm{NL}}~
\label{scalar_nonl_3d}
\eeq
where,
%%%
\beq
\phi^R_{\textrm{NL}}&=&\,\sum_{\tom=0}^{\infty}\sum_{k_{\bot}=-\infty}^{\infty}\f{1}{\sqrt{2\,{\rm Sinh}\,(\f{\pi \tom}{a})}}\,\bigg[\f{d^{1}_{k}}{f(l^2\,\tom^2/a^2,\, l^2(\tom^2-e^{2 a \xi}k_{\bot}^{2}),\,l^2\,k_{\bot}^{2})}\,e^{\f{\pi\,\tom}{2\,a}}\,^{R}u_{\tom k_{\bot}}(\eta,\xi,x_{\bot})\nonumber\\
&+&\,\f{d^{2}_{k}}{f(l^2\,\tom^2/a^2,\, l^2(\tom^2-e^{2 a \xi}k_{\bot}^{2}),\,l^2\,k_{\bot}^{2})}\,e^{-\f{\pi\,\tom}{2\,a}}\,^{R}u^{*}_{\tom -k_{\bot}}(\eta,\xi,x_{\bot})\bigg]+\,h.c\nonumber\\
&=&\,\,\sum_{\tom=0}^{\infty}\sum_{k_{\bot}=-\infty}^{\infty}\f{1}{\sqrt{2\,{\rm Sinh}\,(\f{\pi \tom}{a})}}\,\bigg[p^{1}_{k}\,e^{\f{\pi\,\tom}{2\,a}}\,^{R}u_{\tom k_{\bot}}(\eta,\xi,x_{\bot})+\,p^{2}_{k}\,e^{-\f{\pi\,\tom}{2\,a}}\,^{R}u^{*}_{\tom -k_{\bot}}(\eta,\xi,x_{\bot})\bigg]+\,h.c.~;
\label{scalar_nl_3d_r}
\eeq
and
%%%
\beq
\phi^L_{\textrm{NL}}&=&\,\sum_{\tom=0}^{\infty}\sum_{k_{\bot}=-\infty}^{\infty}\f{1}{\sqrt{2\,{\rm Sinh}\,(\f{\pi \tom}{a})}}\,\bigg[\f{d^{1}_{k}}{f(l^2\,\tom^2/a^2,\, l^2(\tom^2-e^{2 a \xi}k_{\bot}^{2}),\,l^2\,k_{\bot}^{2})}\,e^{-\f{\pi\,\tom}{2\,a}}\,^{L}u^{*}_{\tom -k_{\bot}}(\eta,\xi,x_{\bot})\nonumber\\
&+&\,\f{d^{2}_{k}}{f(l^2\,\tom^2/a^2,\, l^2(\tom^2-e^{2 a \xi}k_{\bot}^{2}),\,l^2\,k_{\bot}^{2})}\,e^{\f{\pi\,\tom}{2\,a}}\,^{L}u_{\tom k_{\bot}}(\eta,\xi,x_{\bot})\bigg]+\,h.c\nonumber\\
&=&\,\,\sum_{\tom=0}^{\infty}\sum_{k_{\bot}=-\infty}^{\infty}\f{1}{\sqrt{2\,{\rm Sinh}\,(\f{\pi \tom}{a})}}\,\bigg[p^{1}_{k}\,e^{-\f{\pi\,\tom}{2\,a}}\,^{L}u^{*}_{\tom -k_{\bot}}(\eta,\xi,x_{\bot})+\,p^{2}_{k}\,e^{\f{\pi\,\tom}{2\,a}}\,^{L}u_{\tom k_{\bot}}(\eta,\xi,x_{\bot})\bigg]+\,H.c.~.
\label{scalar_nl_3d_l}
\eeq
%%%5
Here $\phi^R_{\textrm{NL}}$ $(\phi^L_{\textrm{NL}})$ depicts the field solutions with respect to the Rindler observer who is moving only in the RRW (LRW). 
%%%
One can check that $\phi^R_{\textrm{NL}}$ is exactly the field which is decomposed with respect to right Rindler modes and creation, annihilation operators. 
%%%
%Above we used the solution for Rindler mode as shown in Eq. (\ref{rindler_3d_1}). 
%%%
%To  achieve the final form of the field solutions in LNI-NLT, we take, 
%%%%%
We also have
\beq
p^{1,2}_{k}=\,\f{d^{1,2}_{k}}{f(l^2\,\tom^2/a^2,\, l^2(\tom^2-e^{2 a \xi}k_{\bot}^{2}),\,l^2\,k_{\bot}^{2})}~,
\label{unrun_nonlocal_1}
\eeq
%%%%%
%%%%%%
where $p^{1,2}_{k}$ are interpreted as the nonlocal Unruh operator. As the operator $d^{(1,\,2)}$ annihilates Minkowski vacuum thus $p^{1}_{k}\ket{0_M}=p^{2}_{k}\ket{0_M}=\,0$. Similarly the relations between the local and nonlocal Rindler operators i.e. ${}^{L,R}b_k$ and ${}^{L,R}c_k$ can be found (see Appendix \ref{mod_rindler}). 
The relations between ${}^{L,R}c_k$ and $p^{1,2}_k$ 
can be obtained 
by using the relations between $d^{1,2}_k$, $p^{1,2}_k$ and Eqs (\ref{BRM2}), (\ref{rindler_unruh_rrw}). Therefore one obtains 
\beq
{}^{L}c_{k}=\,\f{1}{\sqrt{2\,{\rm Sinh}\,(\f{\pi \tom}{a})}}\,\bigg[e^{\f{\pi\,\tom}{2\,a}}\,p_{k}^{2}+\,e^{-\f{\pi\,\tom}{2\,a}}\,p^{1^{\dagger}}_{-k}\bigg]~;
\label{BRM3}
\\
{}^{R}c_{k}=\,\f{1}{\sqrt{2\,{\rm Sinh}\,(\f{\pi \tom}{a})}}\,\bigg[e^{\f{\pi\,\tom}{2\,a}}\,p_{k}^{1}+\,e^{-\f{\pi\,\tom}{2\,a}}\,p^{2^{\dagger}}_{-k}\bigg]~.
\label{BRM4}
\eeq
%%%%%
%The commutation relations of the local operators and the above equations provide the commutation relation for the nonlocal theory.
We write the commutation relations for the local operators as follows (see Eq. (2.5.47a) of \cite{takagi}): 
\beq
[{}^{R}b_{k},\,{}^{R}b^{\dagger}_{k'}]=[{}^{L}b_{k},\,{}^{L}b^{\dagger}_{k'}]=\,\d(\tom-\tom')\,\d^2(k_\bot-k^{'}_{\bot})=[d^{1}_{k},\,d^{1^\dagger}_{k'}] = [d^{2}_{k},\,d^{2^\dagger}_{k'}]~,
\label{rin_un_local_com}
\eeq
while the others vanish. Equation (\ref{rin_un_local_com}) and the relations between the local and nonlocal operators govern the nonvanishing commutation relations for the nonlocal Unruh and Rindler operators ({\it i.e} $p$ and $c$) as, 
\beq
[{}^{R}c_{k},\,{}^{R}c^{\dagger}_{k'}] = [{}^{L}c_{k},\,{}^{L}c^{\dagger}_{k'}]&=&\,\f{\d(\tom_k-\tom_k')\,\d^2(k_\bot-k^{'}_{\bot})}{f(l^2\,\tom^{2}/a^2,\, l^2(\tom^{2}-e^{2 a \xi}k_{\bot}^{2}),\,l^2\,k_{\bot}^{2})\,f(l^2\,\tom^{2}/a^2,\, l^2(\tom^{2}-e^{2 a \xi}k_{\bot}^{'2}),\,l^2\,k_{\bot}^{'2})}
%&=&\,\f{\d(\tom-\tom')\,\d^2(k_\bot-k^{'}_{\bot})}{f(k)\,f(k')}
\nonumber
\\
&=& [p^{1}_{k},\,p^{1^\dagger}_{k'}] = [p^{2}_{k},\,p^{2^\dagger}_{k'}]~.
\label{commut_3d_1}
\eeq
We close this section by mentioning the relevant nonlocal forms of the positive frequency Wightman functions with respect to the Rindler proper time in $(1+3)$ dimensions. The Wightman functions in the RRW and LRW can be obtained by using the Eqs. (\ref{scalar_nl_3d_r}) and (\ref{scalar_nl_3d_l}) respectively as, 
\beq
{}^{R,L}G_{W}(\D \t)=\,\f{1}{8 \pi^4 a}\int_{0}^{\infty} d\tom\,\int\,\f{d^2k_{\bot}\,\bigg[K_{\f{i\tom}{a}}\bigg(\f{|k_\bot |}{a}\bigg)\bigg]^2\,\bigg(e^{\f{\pi\,\tom}{a}}\,e^{-i\tom \D \t}+e^{-\f{\pi\,\tom}{a}}\,e^{i\tom \D \t}\bigg)}{f^2\bigg[\f{l^2\tom^{2}}{a^2},\,l^2(\tom^{2}-k_{\bot}^{2}),\,l^2 k_{\bot}^{2}\bigg]}~.
\label{wight_3d_2}
\eeq
For detailed derivation see Appendix  \ref{nl_wight_2d} for $(1+1)$ dimensions and Appendix \ref{nl_wight_3d} for $(1+3)$ dimensions (the same can also be obtained through Rindler quantization, see Appendix \ref{wight_3d_nonlocal}). Note that ${}^{R,L}G_{W}$ is only the function of $\Delta \tau$. Hence it  preserves time translational invariance and respects thermal equilibrium condition. 
However in LNI-NLT this would not be the case if one uses the nonlocal Minkowski quantization and puts Rindler trajectory to obtain the Wightman function in the detector's proper frame \cite{Gim:2018rcy}. 
\section{Unruh effect in $(1+3)$ dimensions}
Having all the prerequisites we are now in a position to investigate the UF effect in LNI-NLT.  We concentrate on two standard approaches: determining the expectation value of Rindler number operator with respect to the Minkowski vacuum and detector's response function. 
In these analyses we assume Unruh quantization of the field belongs to a particular Rindler wedge. 
\subsection{The expectation value of Rindler number operator}
\label{number_3d}
In this section we analyze the expectation value of the nonlocal Rindler number operator with respect to the Minkowski vacuum state.  
Thus we write
\beq
N_k&=&\,\sum_{k'}\bra{0_M}{}^{R}c_{k}^{^{\dagger}}\,{}^{R}c_{k'}\ket{0_M}=\,\mathlarger{\sum}_{k'}\,\f{e^{-\f{\pi(\tom+\tom')}{2 a}}\bra{0_M}p^{2}_{-k}p^{2^\dagger}_{-k'}\ket{0_M}}{\sqrt{4\,{\rm Sinh}\,(\f{\pi \tom}{a})\,{\rm Sinh}\,(\f{\pi \tom'}{a})}}~.
\label{nonlocal_number_unruh_1}
\eeq
%%%
Here we replace $c_k$ in terms of $p_k$ while 
using Eq. (\ref{BRM4}) .
Using the commutation relation of the operator $p$'s from Eq. (\ref{commut_3d_1}) and taking $k= k'$ in Eq. (\ref{nonlocal_number_unruh_1}), we get $N_k$ as follows: 
%%%%
\beq
N_k&=&\,\bra{0_M}{}^{R}c_{k}^{^{\dagger}}\,{}^{R}c_{k}\ket{0_M}=\f{1}{f^2\bigg[(\f{l^2\tom^{2}}{a^2}),\,l^2(\tom^{2}-e^{2 a \xi}\,k_{\bot}^{2})),\,l^2 k_{\bot}^{2}\bigg]}\,\f{1}{(e^{2\pi \tom/a}-1)} \delta(0)~.
\label{nonlocal_number_2}
\eeq
This result exhibits the particle distribution in the Minkowski vacuum as observed by the Rindler observer.
 The appearance of $\delta(0)$ is not new as the same also pops up for local theory. This is merely an artifact of the use of plane wave as basis modes which is nonsquare integrable \cite{Carroll:2004st}. 
Notably, the number operator in Eq. (\ref{nonlocal_number_2}) is modified  
 by an overall multiplicative factor, which bears the nonlocality of the theory. 
On the contrary the temperature of the thermal bath remains unaffected and is given by Unruh temperature $a/2\pi$.
%%%%

Before closing this section we make couple of comments as follows. Here we obtain the expectation value of Rindler number operator
by writing nonlocal Rindler operators in terms of nonlocal Unruh operators. 
The same can also be found by writing ${}^Rc_k$ in terms of nonlocal Minkowski operators. This can be realized by expressing ${}^Rc_k$ in terms of its local counter parts ${}^Rb_k$ and then transforming to the local Minkowski operator ($a_k$) by using the standard Bogoluibov transformation. 
It is well known that use of the  latter relations yields the following result \cite{Crispino,takagi}
\beq
\bra{0_M}{}^{R}b_{k}^{^{\dagger}}\,{}^{R}b_{k}\ket{0_M}=\, \f{1}{(e^{2\pi \tom/a}-1)}\delta(0)~.
\label{number_op_standard}
\eeq
Then use of it in
\begin{equation}
N_k=\,\bra{0_M}{}^{R}c_{k}^{^{\dagger}}\,{}^{R}c_{k}\ket{0_M} = \f{\bra{0_M}{}^{R}b_{k}^{^{\dagger}}\,{}^{R}b_{k}\ket{0_M}}{f^2\bigg[(\f{l^2\tom^{2}}{a^2}),\,l^2(\tom^{2}-e^{2 a \xi}\,k_{\bot}^{2})),\,l^2 k_{\bot}^{2}\bigg]}\,~,
\end{equation}
gives us the required result {\it i.e} Eq. (\ref{nonlocal_number_2}). 
The above outcome is in contradiction with the response function of the UD detector as reported in \cite{Gim:2018rcy}.
 In \cite{Gim:2018rcy}, the response function gets modified nontrivially and the temperature of the thermal bath turns out to be different than $a/2\pi$. 
Below, we show that such contradiction can be avoided when the nonlocal Wightman function is derived using the Unruh quantization. 
\subsection{Response function of the UD detector}\label{response_3d_nonlocal}
%%%%
In this subsection we study the response function of the UD detector where the detector is interacting with a massless scalar field in the background of LNI-NLT. 
%\textcolor{red}{We refer our readers to appendix (\ref{response_2d_nonlocal}) for a brief introduction of the standard response function of the detector corresponding to a local QFT. }
%%%
We put Eq. (\ref{wight_3d_2}) for the RRW into Eq. (\ref{power_1}) and obtain
\beq
I^{R} (\D E)=\,\f{1}{8 \pi^4 a}\int_{0}^{\infty} d\tom\int \f{d^2k_{\bot}\bigg[K_{\f{i\tom}{a}}\bigg(\f{|k_\bot |}{a}\bigg)\bigg]^2}{f^2\bigg[\f{l^2\tom^{2}}{a^2},\,l^2(\tom^{2}-k_{\bot}^{2}),\,l^2 k_{\bot}^{2}\bigg]}\int_{-\infty}^{\infty}d (\D \t)\bigg(e^{\f{\pi\,\tom}{a}}e^{-i(\D E +\tom )\D \t}+e^{-\f{\pi\,\tom}{a}}\,e^{-i(\D E-\tom) \D \t}\bigg)~.
\eeq
%%%
Performing the $\D \t$ and $\tom$ integrals respectively one gets, 
\beq
I^{R} (\D E)&=&\,\f{1}{4 \pi^3 a}\int_{0}^{\infty} d\tom\int \f{d^2k_{\bot}\bigg[K_{\f{i\tom}{a}}\bigg(\f{|k_\bot |}{a}\bigg)\bigg]^2}{f^2\bigg[\f{l^2\tom^{2}}{a^2},\,l^2(\tom^{2}-k_{\bot}^{2}),\,l^2 k_{\bot}^{2}\bigg]}\,\bigg[e^{\f{\pi\,\tom}{a}}\,\d(\D E+\tom)+e^{-\f{\pi\,\tom}{a}}\,\d(\D E-\tom)\bigg]\nonumber\\
&=&\,\f{e^{-\f{\pi\,\D E}{a}}}{4 \pi^3 a}\,\int_0^{2\pi} d\Phi\int_{0}^{\infty}dk_{\bot}\,\f{k_{\bot}\,\bigg[K_{\f{i\D E}{a}}\bigg(\f{|k_\bot |}{a}\bigg)\bigg]^2}{f^2\bigg[\f{l^2\D E^{2}}{a^2},\,l^2(\D E^{2}-k_{\bot}^{2}),\,l^2 k_{\bot}^{2}\bigg]}
\nonumber
\\
&=&\,\f{e^{-\f{\pi\,\D E}{a}}}{2\pi^2 a}\, \int_{0}^{\infty}dk_{\bot}\,\f{k_{\bot}\,\bigg[K_{\f{i\D E}{a}}\bigg(\f{|k_\bot |}{a}\bigg)\bigg]^2}{f^2\bigg[\f{l^2\D E^{2}}{a^2},\,l^2(\D E^{2}-k_{\bot}^{2}),\,l^2 k_{\bot}^{2}\bigg]}~,
\label{res_nl_3d}
\eeq
where in the second equality only $\delta(\Delta E - \tilde{\omega})$ has contributed as $\Delta E>0$. Also $d^2k_{\perp} = k_\perp dk_{\perp} d\Phi$ has been used in the above integral. 
At this stage a couple of comments are in order. 
%%%
\begin{itemize}
\item[i.]
To perform the integral over $k_{\bot}$ in Eq. (\ref{res_nl_3d}), the exact form of the function  $f^2\bigg[\f{l^2\D E^{2}}{a^2},\,l^2(\D E^{2}-k_{\bot}^{2}),\,l^2 k_{\bot}^{2}\bigg]$ need to be known.
\item[ii.] For a local field theory $f=1$ and the integration over $k_{\perp}$ yields $\pi a\Delta E/(2\sinh(\pi\Delta E/a))$. This eventually leads to the standard expression for the response function of the detector. 
\end{itemize}
Note that without choosing a specific form of $f\bigg[\f{l^2\tom^{2}}{a^2},\,l^2(\tom^{2}-k_{\bot}^{2}),\,l^2 k_{\bot}^{2}\bigg]$, the integration in Eq. (\ref{res_nl_3d}) cannot be obtained and therefore the structural similarity between Eqs.(\ref{nonlocal_number_2}) and (\ref{res_nl_3d}) is not apparent. 
 However like the local theory, where these two methods yield similar results, the $(1+1)$ dimensional LNI-NLT produces identical results out of these two methods (compare Eqs. (\ref{BRM22}) and (\ref{BRM222})). 
This cannot be achieved while using the Minkowski quantization (e.g. see \cite{Gim:2018rcy}). This solely happens due to the use of Unruh quantization in the present analysis.  
In the next section we explore the thermality of the Minkowski vacuum state and find the temperature of the thermal bath with respect to the Rindler observer in LNI-NLT.   
\section{Detailed balance form of KMS condition : obtaining the bath temperature}\label{detailed balance_1}
%%%
It can be noted from Eqs (\ref{wight_2dr_1_proper}), (\ref{wight_2dl_1_proper}) and (\ref{wight_3d_2}), that $(1+1)$ and $(1+3)$ dimensional nonlocal Wightman functions exhibit time translational invariance 
with respect to the Rindler proper time. This implies that the thermal equilibrium remains intact in LNI-NLT.  
It is straightforward to show that like the thermal Wightman function at the equilibrium condition, 
these Wightman functions [see Eqs (\ref{wight_2dr_1}), (\ref{wight_2dl_1}), (\ref{wight_3d_2})] satisfy the KMS condition as
%%%
\beq
G_{W} (\D \t-i \,\b)=\,G_{W}(-\D \t)~,
\label{kms_1}
\eeq
%%%
where $\beta = 2\pi/a$ is the inverse temperature of the thermal bath as perceived by the Rindler observer. This temperature is in agreement with our findings in Sec. \ref{number_3d}.
%%%%

In the standard notion of UD detector the ratio of the excitation to deexcitation transition probabilities of the detector turns out to be $e^{-\D E/T}$, where $T$ is the temperature of the thermal bath. 
This relation is a manifestation of the detailed balance form of the KMS condition \cite{Fewster:2016ewy,Garay:2016cpf}, which we mention next. 
%%%%
It can be shown that the Fourier transformed functions of Eq. (\ref{kms_1}) are connected by a relation such as, 
%%%
\beq
\widetilde{G}_{W} (- E)=\,e^{\b  E} G_{W} (E)~,
\label{kms_2}
\eeq
%%%
where
%%%
\beq
\widetilde{G}_{W} (E)=\,\int_{-\infty}^{\infty} d\t \,e^{-i E \t} \,\,G_{W} (\tau)~.
\label{fourier_nl_1}
\eeq
%%%%
The relation (\ref{kms_2}) is known as the detailed balance form of the KMS condition \cite{Fewster:2016ewy,Garay:2016cpf}. 
%%%%
%%%
In addition to the fulfilment of the KMS condition,
we show that the nonlocal Wightman function and the ratio of the transition probabilities of the detector satisfy the detailed balance form of the KMS condition.

We restrict the following analysis in the RRW. It is straightforward to obtain the same for LRW.
Keeping the symbols consistent, we replace $E$ to be $\D E$ and $\t=\D \t$ in Eq. (\ref{fourier_nl_1}), which yields
\beq
\widetilde{G}_{W} (\D E)=\,\int_{-\infty}^{\infty} d(\D \t)\,e^{-i \D E \D \t}\, G_{W} (\D \t)~.
\label{detailed_kms}
\eeq
This equation is identical to Eq. (\ref{power_1}). Therefore Eq. (\ref{detailed_kms}) becomes
\beq
\widetilde{G}_{W} (\D E)
=\,\f{e^{-\f{\pi \D E}{a}}}{4\, \pi a^3}\,\int\,\f{d^2k_{\bot}\bigg[K_{\f{i\D E}{a}}\bigg(\f{|k_\bot |}{a}\bigg)\bigg]^2}{f^2\bigg[\f{l^2\D E^{2}}{a^2},\,l^2(\D E^{2}-k_{\bot}^{2}),\,l^2 k_{\bot}^{2}\bigg]}~.
\label{fourier_1}
\eeq
%%%
Replacing $\D E \to -\,\D E $ in the above equation we obtain, 
\beq
\widetilde{G}_{W} (- \D E)=\,\f{e^{\f{\pi \D E}{a}}}{2\, \pi a^3}\,\int\, \f{d^2k_{\bot}\bigg[K_{\f{i\D E}{a}}\bigg(\f{|k_\bot |}{a}\bigg)\bigg]^2}{f^2\bigg[\f{l^2\D E^{2}}{a^2},\,l^2(\D E^{2}-k_{\bot}^{2}),\,l^2 k_{\bot}^{2}\bigg]}~.
\label{fourier_2}
\eeq
%%
%%%
Thus the ratio of the Eqs. (\ref{fourier_1}) and (\ref{fourier_2}) turns out to be 
%%%%%
\beq
\widetilde{G}_{W} (- \D E)=\,e^{\b \D E}\,\widetilde{G}_{W} (\D E)~,
\label{ratio_3d_1}
\eeq
where $\b =\, 2\pi/a$. Therefore the $(1+3)$ dimensional nonlocal Wightman function satisfies the detailed balance form along with the KMS condition. Furthermore it is straightforward to show that the ratio ($\mathcal{R}$) of the excitation to deexcitation transition probabilities of the detector  [calculated from Eq.(\ref{res_nl_3d})] reduces to 
%%%
\beq
\mathcal{R} (\D E,\,s)\,\,\,\xrightarrow[s \to 0]\,\,\,\f{I^R (\D E)}{I^R (- \D E)}\,\,=\,\,\,e^{- \b\,\D E}~.
\label{ratio_3d_2}
\eeq
This implies that in LNI-NLT, the temperature of thermal bath comes out to be $T_{\rm NL}=\f{a}{2\pi}$,  which is the same as the standard Unruh temperature.
In contrast to \cite{Gim:2018rcy}, our results suggest that the temperature of the thermal bath remains unchanged even in the Lorentz violating nonlocal theory. 
%%%
Proceeding similarly, one can show that the above result holds in
$(1+1)$ dimensional LNI-NLT. 
\section{Lorentz invariant model using Unruh mode}\label{LI}
%%%
In this section we discuss the use of Unruh quantization in the LI-NLT. In particular we discuss the LI-NLT models as considered in \cite{Kajuri:2017jmy,Modesto:2017ycz}. In this class of theories the equation of motion for a quantum scalar field obeys a homogeneous and infinite order differential equation such as 
\begin{equation}
\Box\Big(l^2F(\Box)\Big)\phi=0~,
\label{BRM4n} 
\end{equation}
%%%%
where $F$ is an analytical function and possesses nonzero value everywhere. 
The solution to Eq. (\ref{BRM4})
is well developed in \cite{Barnaby:2007ve,Kajuri:2017jmy,Modesto:2017ycz}. Their results describe that the number of solutions is equal to the number of poles present in its propagator, which turns out to be $G(p^2)=\,\frac{F(-l^2p^2)^{-1}}{p^2}$ \cite{Barnaby:2007ve,Kajuri:2017jmy}.
Thus Eq. (\ref{BRM4n}) owns two independent solutions. It can be intuitively perceived that the solutions to the equation  $\Box \phi=\,0$ satisfy Eq. (\ref{BRM4n}). Note that Eq. (\ref{BRM4n}) has only two solutions and therefore it implies that the complete set of solutions to Eq. (\ref{BRM4n}) is same as that of the  $\Box \phi=\,0$. Therefore, like LNI-NLT, the modes in this case are the same as local theory. Hence following the analysis for LNI-NLT, the field solutions in LI-NLT can be written in terms of the nonlocal Unruh operators [e.g. (\ref{scalar_nl_3d_r}) and (\ref{scalar_nl_3d_l})]. However now the nonlocal operators are related to local counterparts by the following relation:
\begin{equation}
p^{1,2}_{k}=\,\f{d^{1,2}_{k}}{F(-l^2k_\mu k^\mu)}~.
\label{LIU}
\end{equation}
Since these calculations are done under the on-shell condition for a massless scalar field i.e. $k_\mu k^\mu=0$, the denominator turns out to be $F(0)$ in the above equation. 
%%%
Note that such is also true for $l=0$, which is recognized as the local field limit. We already have $F(0) = 1$ for $l=0$; therefore it suggests  that in all situations (local and nonlocal) one must have $F(0)=1$. This implies that all the nonlocal operators are identical to their local versions and hence 
the UF effect via Bogoliubov coefficient method \cite{Kajuri:2017jmy} and UD detector method \cite{Modesto:2017ycz} (irrespective of the choice of mode solutions) stands unaltered in LI-NLT. 
%%%
%%
On the contrary, LNI-NLT acquires modifications in the UF effect due to the violation of Lorentz invariance, still the temperature of the thermal bath remains unaltered.
\section{Discussions}\label{discussions}
%%%%
Studying UF effect in the context of nonlocal field theories is an arguable and gray area of research \cite{Nicolini:2009dr,Kajuri:2017jmy,Modesto:2017ycz,Gim:2018rcy}. 
%%%
We do not have a clear perception of which constituents control the outcome of the UF effect in nonlocal theories. In this work we aim to examine the UF effect in a class of nonlocal theories as considered in \cite {Gim:2018rcy}. 
Contrary to \cite{Gim:2018rcy}, we write the field solutions using Unruh quantization and study its impacts in the UF effect. 
%
%We derive the nonlocal Wightman functions and examine the 
We follow both the number operator and detector methods in order to examine the nonlocal UF effect. 
We summarize our findings in the following order :
(i) In LNI-NLT, the Wightman functions in both dimensions differ from the local one by a form factor, which involves the minimal length scale associated with the theory.  As a consequence the response function of the UD detector acquires modifications in the present work. This is a different outcome than was predicted in \cite{Modesto:2017ycz} in the context of LI-NLT.  We also show that the expectation values of the Rindler number operators get modified by an overall form factor.
%%%
(ii) In our work the nonlocal Wightman functions remain time translationally invariant with respect to the Rindler proper time. This signifies that the detector 
remains in thermal equilibrium with the scalar field. Furthermore we show that the temperature of the thermal bath retains its standard form as a  local UF effect.  
%%%
These results contradict the findings in \cite {Gim:2018rcy} which predict a time translationally noninvariant Wightman function and modification in the temperature of the thermal bath.
(iii) In our case the nonlocal Wightman functions and the ratio between the excitation to deexcitaion probabilities of the UD detector satisfy the detailed balance form of the KMS condition. These strengthens the speculation that the detector detects the temperature of the thermal bath as $T_{\rm{NL}}= a/2\pi$. This differs 
from the temperature as reported in \cite {Gim:2018rcy}. In our analysis the use of Unruh quantization plays the pivotal
role which guarantees thermal equilibrium and thereby provides an unambiguous
 definition of temperature.  

The present work can be categorized as a survey where we 
review the controversies regarding the nonlocal UF 
effect and bring in a new line of thought, whether 
the nonlocal UF effect depends on 
the choice of a particular quantization of the field. 
Our work demonstrates that different quantization mechanisms yield nonidentical results in the context of nonlocal field theories. It also portrays that choice of field quantization plays a significant role in determining the thermal nature of a system and the outcome of nonlocal UF effect.
We do not claim that our findings may resolve the long-standing debate, rather we comment that studies related to the nonlocal UF effect cannot be fit under a single umbrella. 
In this regard, recognizing the intriguing yet diverse results, one should continue to investigate the effects of all the constituent parameters of possible nonlocal models. It would be important to investigate the nonlocal effects in the context of quantum entanglement, entanglement harvesting between the two uniformly accelerating detectors within the present setup. The work is under progress.
%%%
%%%%%
%%%%%%%%%%%%%%%%%%
\section*{ACKNOWLEDGMENT}
%%%%%%%%%%%%%%%%%%
A.D. would like to thank the Indian Association for the Cultivation of Science, Kolkata, for providing an academic visit, during which the present work was formulated.

%%%%
\appendix 
\section*{Appendices}
%%%%
%%%%%
\section{NONLOCAL
SCALAR
FIELD
IN $(1+3)$ DIMENSIONAL MINKOWSKI SPACETIME}\label{mod_minkowski}
%%%%
%%%
%In literature \cite{Barnaby:2010kx,Gim:2018rcy}, it is well discussed that the number of independent solutions for an infinite order differential equation (e.g. Eq. (\ref{scalar_nonl_1})) depend on the number of poles that appear in its propagator. As the number of poles in the propagator turns out to be same as one obtains in the local quantum field equation, the field solutions for the nonlocal theory also appear to be the same set of solutions which one gets by solving $\Box\,\phi=\,0$. 
%%%%
%Thus, in the Minkowski spacetime, the plane wave solutions $\sim e^{\pm ik_{\mu}x^\mu}$  can be considered as the set of solutions for the nonlocal field equation as depicted in Eq. (\ref{scalar_nonl_1}). Here $k_{\mu}=({\o,\,\vec{k}})$. 
%%%%%
%Note that the nonlocal effects of the background theory make an appearance through the creation and the annihilation operators which leads to the modifications in the operator sector instead the mode solutions. As a consequence the commutation relations between the operators also get altered than that of the local one due to the nonlocal effects \cite{Gim:2018rcy}. 
%%
%
Following the discussion in the Sec. \ref{model} we write the nonlocal field solutions in terms of the Minkowski modes as follows: 
\beq
\phi_{{\rm NL}}(x)\,=\,\int \f{d^3k}{(2\pi)^3\,2\o}\,\bigg[g_{k}\,e^{ik_{\mu}x^\mu}+\,g_{k}^{\dagger}\,e^{-ik_{\mu}x^\mu}\bigg]~.
\label{min_nonl_1}
\eeq
%%\
Here $(g_k, g_{k}^{\dagger})$ depict the nonlocal version of the local Minkowski operators which we symbolize as ($a_k,\, a_{k}^{\dagger}$). 
It is worth mentioning that the mode decomposition for the local field, satisfying, equation 
$\Box\,\phi=\,0$ is taken as below :
%%%
%
\beq
\phi(x)\,=\,\int \f{d^3k}{(2\pi)^3\,2\o}\,\bigg[a_{k}\,e^{ik_{\mu}x^\mu}+\,a_{k}^{\dagger}\,e^{-ik_{\mu}x^\mu}\bigg]~,
\label{scalar_loc_1}
\eeq
where the operators satisfy the commutation relations $[a_k,\,a_{k'}^{\dagger}]=\,(2\pi)^3\,2\o\,\d^3(\vec{k}-\vec{k'})$ and $[a_k,\,a_{k'}]=[a_{k}^{\dagger},\,a_{k'}^{\dagger}]=0$.
%%
%%%%%
Using Eqs. (\ref{scalar_loc_nl}) and (\ref{scalar_loc_1}) one can perceive the relation between the local and the nonlocal Minkowski operators as \cite{Gim:2018rcy} 
%%%
\beq
\phi_{\rm NL}&=&\,f^{-1}\,((i\,l\,\p_{0})^2,\,(i\,l\,\p_{i})^2)\,\phi\nonumber\\
&=&\,\int \f{d^3k}{(2\pi)^3\,2\o}\,\bigg[\f{a_{k}}{f(l^2\,\o^{2},\, l^2\,k^{2})}\,e^{ik_{\mu}x^\mu}+\,\f{a_{k}^{\dagger}}{f(l^2\,\o^{2},\, l^2\,k^{2})}\,e^{-ik_{\mu}x^\mu}\bigg]~.
\label{scalar_nonl_2}
\eeq
%%%%%
Comparing Eq. (\ref{min_nonl_1}) and Eq. (\ref{scalar_nonl_2}) one gets
%%%%%
\beq
g_{k}=\,\f{a_{k}}{f(l^2\,\o^{2},\, l^2\,k^{2})};\,\,\,\,\,\,\,\,\,\,\,\,\,\,\,\,\,\,\,\,\,g_{k}^{\dagger}=\,\f{a_{k}^{\dagger}}{f(l^2\,\o^{2},\, l^2\,k^{2})}~.
\label{relation_min_1}
\eeq
Furthermore from the standard commutation relation of local Minkowski operators we find their nonlocal counterparts as
%%%%
\beq
[g_k,\,g_{k'}^{\dagger}]=\,\f{(2\pi)^2\,\o\,\d^3(\vec{k}-\vec{k'})}{f\big(l^2\,\o^{2},\, l^2\,k^{2}\big)\,f\big(l^2\,\o^{2},\, l^2\,k^{'2}\big)}; \,\,\,\,\,\,\,\,\,\,\,[g_k,\,g_{k'}]=[g_{k}^{\dagger},\,g_{k'}^{\dagger}]=0~.
\eeq
Equation (\ref{relation_min_1}) dictates that the vacuum state corresponding to the nonlocal field theory is the same as that of the local Minkowski vacuum state and characterized by :  $g_k\ket{0}_M=0$. 
\section{NONLOCAL SCALAR FIELD IN $(1+3)$ \& $(1+1)$ DIMENSIONAL RINDLER SPACETIME USING RINDLER QUANTIZATION}\label{mod_rindler}
%%%
As mentioned above, in Rindler spacetime the mode solutions of the local field will also be those for the nonlocal ones. Therefore in $(1+3)$ dimensions the nonlocal scalar field in Rindler spacetime is decomposed as 
%%%
\beq
\phi_{\rm NL} (x)&=&\,\int_{0}^{\infty}d\tom\int d^{2}k_{\bot}\, [{}^{R}c_{\tom k_{\bot}}\,{}^{R}u_{\tom k_{\bot}}(\eta,\xi,x_{\bot})+{}^{R}c_{\tom k_{\bot}}^{^{\dagger}}\,^{R}u^{*}_{\tom k_{\bot}}(\eta,\xi,x_{\bot})\nonumber\\
&+&\,{}^{L}c_{\tom k_{\bot}}\,^{L}u_{\tom k_{\bot}}(\eta,\xi,x_{\bot})+{}^{L}c_{\tom k_{\bot}}^{^{\dagger}}\,^{L}u^{*}_{\tom k_{\bot}}(\eta,\xi,x_{\bot})]
\nonumber
\\
&\equiv& \phi_{\rm NL}^R (x) + \phi_{\rm NL}^L (x)~.
\label{field_rind_nl_1}
\eeq
%%%%
Here ${}^{R}c_{\tom k_{\bot}}$ and ${}^{R}c_{\tom k_{\bot}}^{^{\dagger}}$ symbolize the nonlocal counterpart of local Rindler annihilation (${}^{R}b_{\tom k_{\bot}}$) and creation (${}^{L}b_{\tom k_{\bot}}^{^{\dagger}}$) operators, and so on.
The standard local Rindler field is decomposed as follows: 
\beq
\phi (x)&=&\,\int_{0}^{\infty}d\tom \int d^{2}k_{\bot}\, [{}^{R}b_{\tom k_{\bot}}\,{}^{R}u_{\tom k_{\bot}}(\eta,\xi,x_{\bot})+{}^{R}b_{\tom k_{\bot}}^{^{\dagger}}\,^{R}u^{*}_{\tom k_{\bot}}(\eta,\xi,x_{\bot})\nonumber\\
&+&\,{}^{L}b_{\tom k_{\bot}}\,^{L}u_{\tom k_{\bot}}(\eta,\xi,x_{\bot})+{}^{L}b_{\tom k_{\bot}}^{^{\dagger}}\,^{L}u^{*}_{\tom k_{\bot}}(\eta,\xi,x_{\bot})]
\nonumber
\\
&\equiv& \phi^R (x) + \phi^L(x)~.
\eeq
In the above two equations we denote the Rindler mode function for RRW and LRW as : ${}^{R,L}u_{\tom k_{\bot}}(\t,\xi,x_{\bot})$,
which is depicted in Eq. (\ref{rindler_3d_1}). 
%%%
%\beq
%{}^{R}u_{\tom k_{\bot}}(\eta,\xi,x_{\bot})=\,\bigg[\f{{\rm Sinh}(\pi \tom/a)}{4 \pi^4\,a}\bigg]^{1/2}\,K_{\f{i\tom}{a}}\bigg(\f{|k_\bot |\,e^{a\xi}}{a}\bigg)\,e^{i k_\bot.x_{\bot}-\,i\tom \eta}
%\label{rindler_3d_1}
%\eeq
%%
%Note : $\eta=a\,\t$.
%%
All the $b_k$'s are Rindler creation and annihilation operators hence, $b_k \ket{0_R}=\,0$ where $\ket{0_R}$ represents the Rindler vacuum. 
Let us now focus on the nonlocal field solution in the RRW: 
%\beq
%&&\phi_{\rm NL}(x)=\, f^{-1}\,((i\,l\,\p_{0})^2,\,(i\,l\,\p_{i})^2)\,\phi(x)
%\eeq
%%%
\beq
&&\phi_{\rm NL}^{R}(x)=\,f^{-1}\,((i\,l\,\p_{0})^2,\,(i\,l\,\p_{i})^2)\,\int_{0}^{\infty}d\tom \int d^{2}k_{\bot}\, [{}^{R}b_{\tom k_{\bot}}\,{}^{R}u_{\tom k_{\bot}}(\eta,\xi,x_{\bot})+{}^{R}b_{\tom k_{\bot}}^{^{\dagger}}\,^{R}u^{*}_{\tom k_{\bot}}(\eta,\xi,x_{\bot})]\nonumber\\
&&=\int_{0}^{\infty}d\tom \int d^{2}k_{\bot}\,\bigg[\f{{}^{R}b_{\tom k_{\bot}}\,{}^{R}u_{\tom k_{\bot}}(\eta,\xi,x_{\bot})}{f\big((\f{l^2\tom^{2}}{a^2}),\,(l^2(\tom^{2}-e^{2 a \xi}\,k_{\bot}^{2})),\,l^2 k_{\bot}^{2}\big)}+\f{{}^{R}b_{\tom k_{\bot}}^{^{\dagger}}\,^{R}u^{*}_{\tom k_{\bot}}(\eta,\xi,x_{\bot})}{f\big((\f{l^2\tom^{2}}{a^2}),\,(l^2(\tom^{2}-e^{2 a \xi}\,k_{\bot}^{2})),\,l^2 k_{\bot}^{2}\big)}\bigg]~.
\eeq
%%%%%
Therefore following the earlier argument we get
%%%
\beq
{}^{R}c_{k}=\, \f{{}^{R}b_{\tom k_{\bot}}}{f\big((\f{l^2\tom^{2}}{a^2}),\,(l^2(\tom^{2}-e^{2 a \xi}\,k_{\bot}^{2})),\,l^2 k_{\bot}^{2}\big)};\,\,\,\,\,\,\,\,\,\,\,\,\,\,\,\,\,\,\,\,\,\,\,\,{}^{R}c_{k}^{^{\dagger}}=\,\f{{}^{R}b_{\tom k_{\bot}}^{^{\dagger}}}{f\big((\f{l^2\tom^{2}}{a^2}),\,(l^2(\tom^{2}-e^{2 a \xi}\,k_{\bot}^{2})),\,l^2 k_{\bot}^{2}\big)}~.
\label{BRM1}
\eeq
%%%%
Using the standard commutation relation of ${}^{R}b_{\tom k_{\bot}}$ and ${}^{R}b_{\tom k_{\bot}}^{^{\dagger}}$, the commutator of nonlocal Rindler operators becomes  
\beq
\bigg[{}^{R}c_{k},\,{}^{R}c_{k'}^{^{\dagger}}\bigg]=\,\f{\d(\tom - \tom')\,\d^2(k_\bot - k'_{\bot})}{f\big((\f{l^2\tom^{2}}{a^2}),\,(l^2(\tom^{2}-e^{2 a \xi}\,k_{\bot}^{2})),\,l^2 k_{\bot}^{2}\big)\,f\big((\f{l^2\tom'^{2}}{a^2}),\,(l^2(\tom'^{2}-e^{2 a \xi}\,k_{\bot}^{'2})),\,l^2 k_{\bot}^{'2}\big)}~.
\eeq
%%%%

Proceeding similarly in case of ($1+1$) dimensional spacetime one can write, 
\beq
&&\phi_{\rm NL}^{R}(x)=\,f^{-1}\,((i\,l\,\p_{0})^2,\,(i\,l\,\p_{i})^2)\,\int_{-\infty}^{\infty}dk \, [^{R}b_{k}\,{}^{R}u_{k}(\eta,\xi)+^{R}b_{k}^{\dagger}\,\,{}^{R}u_{k}^{*}(\eta,\xi)]~,
\eeq
%%%
where the local Rindler modes are given by \cite{birrell}
\beq
{}^{R}u_{k}&=&\,\f{1}{\sqrt{4\pi\,\tom}}\,e^{ik\xi-\,i\tom\,\eta}\quad \quad  (\rm{in \,\,RRW})\nonumber\\
\nonumber\\
&=&\quad 0 \quad \,\,\,\,\,\,\,\,\,\,(\rm{in \,\,LRW})~;
\label{nonl_rrw_2}
\eeq
and
\beq
{}^{L}u_{k}&=&\,\f{1}{\sqrt{4\pi\,\tom}}\,e^{ik\xi+\,i\tom\,\eta}\quad \quad  (\rm{in \,\,LRW})\nonumber\\
\nonumber\\
&=&\quad 0 \quad \,\,\,\,\,\,\,\,\,\,(\rm{in \,\,RRW})~.
\label{nonl_lrw_2}
\eeq
%%% 
We mention that in $(1+1)$ dimensional massless case we have $\tom=\,|k|$, $\tom >\,0$. Therefore one finds
%%%
\beq
\phi_{\rm NL}^{R}(x)=\,\int_{-\infty}^{\infty}d k\, \bigg[\,\f{^{R}b_{k}\,{}^{R}u_{k}\,(\eta,\xi)}{f\big(l^2\tom^{2},\,l^2 k^{2}\big)}+\f{^{R}b_{k}^{\dagger}\,\,{}^{R}u_{k}^{*}\,(\eta,\xi)}{f\big(l^2\tom^{2},\,l^2 k^{2}\big)}\,\bigg]=\,\,\int_{-\infty}^{\infty}d k\, \bigg[\,^{R}c_k\,{}^{R}u_{k}\,(\eta,\xi)+\,^{R}c_{k}^{\dagger}\,\,{}^{R}u_{k}^{*}\,(\eta,\xi)\,\bigg]
\eeq
and thus we obtain the commutation relation between the nonlocal Rindler operators in $1+1$ dimensions as follows, 
%%%%
\beq
\bigg[{}^{R}c_{k},\,{}^{R}c_{k'}^{^{\dagger}}\bigg]=\,\f{\d(k - k')}{f\big(l^2 \tom^2,\,l^2 k^{2}\big)\,f\big(l^2 \tom^{'2},\,l^2 k^{'2}\big)}~.
\eeq
%%
%%%
\section{NONLOCAL UNRUH MODES AND NUMBER OPERATOR IN $(1+1)$ dIMENSIONS}\label{numberop_unmode}
%%%
%%%%%%
%%%%
Proceeding similarly, as has been done for $(1+3)$ dimensions in Sec. \ref{numberop_unmode1}, in the case of ($1+1$) dimensional spacetime one can write the field in terms of Unruh mode as follows:
%%%%
\beq
\phi(x)=\,\sum_{k=-\infty}^{\infty}\f{1}{\sqrt{2\,{\rm Sinh}\,(\f{\pi \tom}{a})}}\,\bigg[d^{1}_{k}\,(e^{\f{\pi\,\tom}{2\,a}}\,{}^{R}u_{k}+\,e^{-\f{\pi\,\tom}{2\,a}}\,{}^{L}u^{*}_{-k})+\,d^{2}_{k}\,(e^{-\f{\pi\,\tom}{2\,a}}\,{}^{R}u^{*}_{- k}+\,e^{\f{\pi\,\tom}{2\,a}}\,{}^{L}u_{k})\bigg]+\,{\rm H.c}.~.
\eeq
%%
%%%
Subsequently using Eq. (\ref{scalar_loc_nl}) one can write the nonlocal scalar field as
\beq
\phi_{\rm NL}&=&\,\sum_{k=-\infty}^{\infty}\f{1}{\sqrt{2\,{\rm Sinh}\,(\f{\pi \tom}{a})}}\,\bigg[f^{-1}(l^2\,\tom^2,\, l^2\,k^{2})\,d^{1}_{k}\,(e^{\f{\pi\,\tom}{2\,a}}\,{}^{R}u_{k}+\,e^{-\f{\pi\,\tom}{2\,a}}\,{}^{L}u^{*}_{-k})\nonumber\\
&+&\,f^{-1}(l^2\,\tom^2,\, l^2\,k^{2})\,d^{2}_{k}\,(e^{-\f{\pi\,\tom}{2\,a}}\,{}^{R}u^{*}_{- k}+\,e^{\f{\pi\,\tom}{2\,a}}\,{}^{L}u_{k})\bigg]+\,{\rm H.c.}\nonumber\\
&=&\,\sum_{k=-\infty}^{\infty}\f{1}{\sqrt{2\,{\rm Sinh}\,(\f{\pi \tom}{a})}}\,\bigg[p^{1}_{k}\,(e^{\f{\pi\,\tom}{2\,a}}\,{}^{R}u_{k}+\,e^{-\f{\pi\,\tom}{2\,a}}\,{}^{L}u^{*}_{-k})+\,p^{2}_{k}\,(e^{-\f{\pi\,\tom}{2\,a}}\,{}^{R}u^{*}_{- k}+\,e^{\f{\pi\,\tom}{2\,a}}\,{}^{L}u_{k})\bigg]+\,{\rm H.c.}~,
\label{scalar_nonl_3}
\eeq
%%%%%
where,  $p_{k}^{1}=\,f^{-1}(l^2\,\tom^2,\, l^2\,k^{2})\,d^{1}_{k}$. 
From Eq. (\ref{scalar_nonl_3}) we write the mode decomposition in separate wedges as 
%%%%%
\beq
\phi_{\rm NL}^{R}&=&\,\sum_{k=-\infty}^{\infty}\f{1}{\sqrt{2\,{\rm Sinh}\,(\f{\pi \tom}{a})}}\,\bigg[p^{1}_{k}\,e^{\f{\pi\,\tom}{2\,a}}\,{}^{R}u_{k}+\,\,p^{2}_{k}\,e^{-\f{\pi\,\tom}{2\,a}}\,{}^{R}u^{*}_{- k}\bigg]+ {\rm H.c.}~;
\label{nonl_rrw_1}\\
\phi_{\rm NL}^{L}&=&\,\sum_{k=-\infty}^{\infty}\f{1}{\sqrt{2\,{\rm Sinh}\,(\f{\pi \tom}{a})}}\,\bigg[p^{1}_{k}\,e^{-\f{\pi\,\tom}{2\,a}}\,{}^{L}u^{*}_{- k}+\,\,p^{2}_{k}\,e^{\f{\pi\,\tom}{2\,a}}\,{}^{L}u_{k}\bigg]+ {\rm H.c.}~.
\label{nonl_lrw_1}
\eeq
%%%%%%%
%%
%
 Using the commutation relation between the $d$ operators the commutation relation between $p$'s in $(1+1)$ dimensions comes out to be  
%%%%%%
\beq
[\,p_{k}^{1},\,p_{k'}^{1^{\dagger}}\,]=\,\f{\d(k-k')}{f(l^2\tom,\,l^2 k^2)\,f(l^2\tom^{2},\,l^2 k'^2)}~.
\eeq
%%%
The expectation value of the nonlocal Rindler number operator using the Unruh mode in $(1+1)$ dimensions turns out to be as follows.
We proceed similarly as in the $(1+3)$ dimensional case where we  
write the operators $b$ in terms of $p$'s in $(1+1)$ dimensions and obtain the expectation value of the number operator as,
%%
%%%
\beq
N_k&=&\,\bra{0_M}c_{k}^{R^{\dagger}}\,c_{k}^{R}\ket{0_M}=\,\f{1}{f^2\big(l^2 \tom^2,\,l^2 k^{2}\big)}\,\bra{0_M}b_{k}^{R^{\dagger}}\,b_{k}^{R}\ket{0_M}=\,\f{1}{f^2\big(l^2 \tom^2,\,l^2 k^{2}\big)}\,\f{\d(0)}{(e^{2\pi \tom/a}-1)}~.
\label{BRM22}
\eeq
%%%
%%%%%%%%%%
\section{RESPONSE FUNCTION OF The UD DETECTOR IN $(1+1)$ DIMENSIONAL NONLOCAL THEORY}\label{response_2d_nonlocal}
%%%%%

%%%%%%%%%%%%%%%%
\subsection{Nonlocal Wightman functions in $(1+1)$ dimensions}\label{wight_2d_nonlocal}\label{nl_wight_2d}
%%%%%
The positive frequency Wightman function for the observer in the right Rindler wedge is given by
\beq
{}^{R}G_{W}(x_{1},\,x_{2 })=\,\bra{0_M}{\phi}^R_{\rm NL}(x_{1})\,{\phi}^{R}_{\rm NL}(x_{2 })\ket{0_M}~.
\eeq
%%%%%
Using Eq. (\ref{nonl_rrw_1}) the above becomes
%%%%
\beq
{}^{R}G_{W}(x_{1},\,x_{2 })&=&\,\int_{-\infty}^{\infty}\int_{-\infty}^{\infty}\f{dk_{1 }\,dk_{2 }}{2\,\sqrt{{\rm Sinh}\,\left(\f{\pi \tom_1}{a}\right)\,{\rm Sinh}\,\left(\f{\pi \tom_2}{a}\right)}}\bigg[\bra{0_M}p^{1}_{k_{1 }}p^{1^\dagger}_{k_{2 }}\ket{0_M}\,e^{\f{\pi\tom_{1}}{2 a}}\,e^{\f{\pi\tom_{2}}{2 a}}\,{}^{R}u_{k_{1}}(x_{1})\,{}^{R}u^{*}_{k_{2 }}(x_{2})\nonumber\\
&+& \,\bra{0_M}p^{2}_{k_{1}}p^{2^\dagger}_{k_{2 }}\ket{0_M}\,e^{-\f{\pi \tom_{1}}{2 a}}\,e^{-\f{\pi \tom_{2}}{2 a}}\,{}^{R}u^{*}_{- k_{1}}(x_{1})\,{}^{R}u_{- k_{2}}(x_{2 })\bigg]
\nonumber
\\
&=& \,\int_{-\infty}^{\infty}\int_{-\infty}^{\infty}\f{dk_1\,dk_2}{2\,\sqrt{{\rm Sinh}\,\left(\f{\pi \tom_1}{a}\right)\,{\rm Sinh}\,\left(\f{\pi \tom_2}{a}\right)}}\f{\d(k_1-k_2)}{f(l^2\tom_{1}^{2},\,l^2k_{1}^2)\,f(l^2\tom_{2}^{2},\,l^2k_{2}^2)}\nonumber\\
&\times& \bigg(e^{\f{\pi\tom_1}{2 a}}\,e^{\f{\pi\tom_2}{2 a}}\,{}^{R}u_{k_{1}}(x_{1})\,{}^{R}u^{*}_{k_{2 }}(x_{2 })
+ \,e^{-\f{\pi \tom_1}{2 a}}\,e^{-\f{\pi \tom_2}{2 a}}\,{}^{R}u^{*}_{- k_{1}}(x_{1})\,{}^{R}u_{- k_{2}}(x_{2})\bigg)~.
\label{wightman_rrw_1}
\eeq
Integrating over $k_{2}$, we take $k_{1}=\,k_{2}=\,k$. Note that in the massless $(1+1)$ dimensional case we also take $\tom=\,|k|$, $\tom > 0$. Therefore one finds
\beq
{}^{R}G_{W}(x_{1},\,x_{2 })=\,\int_{-\infty}^{\infty}\f{dk}{2\,{\rm Sinh}\left(\f{\pi \tom}{a}\right)}\f{1}{f^2(l^2 \tom^{2},\,l^2 k^2)}\bigg[e^{\f{\pi \tom}{a}}\,{}^{R}u_{k}(\xi_{1},\,\eta_{1}){}^{R}u^{*}_{k}(\xi_{2},\,\eta_{2})
+\,e^{-\f{\pi \tom}{a}}\,{}^{R}u^{*}_{- k}(\xi_{1},\,\eta_{1})\,{}^{R}u_{- k}(\xi_{2},\,\eta_{2})\bigg].
\eeq
Use of the explicit forms of the modes as stated in Eq.(\ref{nonl_rrw_2}) yields 
 %%%%%
\beq
{}^{R}G_{W}(x_{1},\,x_{2})=\,\int_{-\infty}^{\infty}\f{dk}{8\pi\,\tom\,{\rm Sinh}\left(\f{\pi \tom}{a}\right)}\,\f{1}{f^2(l^2 \tom^{2},\,l^2 k^2)}\bigg[e^{\f{\pi \tom}{a}}\,e^{i k \D \xi_{1 2}-\,i \tom \D\eta_{1 2}}
+\,e^{-\f{\pi \tom}{a}}\,e^{i k \D\xi_{1 2}+\,i \tom \D\eta_{1 2}}\bigg]~,
\label{wight_2dr_1}
\eeq
where $\D\xi_{1 2}=\,\xi_{1}-\,\xi_{2}$,  $\D\eta_{1 2}=\,\eta_{1 }-\,\eta_{2}$. Imposing the proper frame condition for the Rindler observer such as $\xi_1=\xi_2=0$ and $\eta_{1,2}=\t_{1,2}$, we obtain
\beq
{}^{R}G_{W}(\D \t)=\,\int_{-\infty}^{\infty}\f{dk}{8\pi\,\tom\,{\rm Sinh}\left(\f{\pi \tom}{a}\right)}\,\f{1}{f^2(l^2 \tom^{2},\,l^2 k^2)}\bigg[e^{\f{\pi \tom}{a}}\,e^{- i \tom \D \t}
+\,e^{-\f{\pi \tom}{a}}\,e^{i \tom \D \t}\bigg]~.
\label{wight_2dr_1_proper}
\eeq
Here $\t_{1,2}$ symbolizes the proper time of the accelerated observer and we take $\t_1-\t_2=\,\D \t$.

Now replacing $\D\eta_{1 2}\to\,-\,\D\eta_{1 2}$ we get the positive frequency Wightman function for the detector in the LRW as follows:
\beq
{}^{L}G_{W}(x_{1},\,x_{2})=\,\int_{-\infty}^{\infty}\f{dk}{8\pi\,\tom\,{\rm Sinh}\left(\f{\pi \tom}{a}\right)}\,\f{1}{f^2(l^2 \tom^{2},\,l^2 k^2)}\bigg[e^{\f{\pi \tom}{a}}\,e^{i k \D \xi_{1 2}+\,i \tom \D\eta_{1 2}}
+\,e^{-\f{\pi \tom}{a}}\,e^{i k \D\xi_{1 2}-\,i \tom \D\eta_{1 2}}\bigg]~.
\label{wight_2dl_1}
\eeq
%%%%
We put the proper frame condition for the Rindler observer in the LRW as $\xi_1=\xi_2=0$ and $\eta_{1,2}=\,-\,\t_{1,2}$ in the above equation, which leads $\D \eta_{12} = - \D \t$ in the LRW. Thus one obtains, 
\beq
{}^{L}G_{W}(\D \t)=\,\int_{-\infty}^{\infty}\f{dk}{8\pi\,\tom\,{\rm Sinh}\left(\f{\pi \tom}{a}\right)}\,\f{1}{f^2(l^2 \tom^{2},\,l^2 k^2)}\bigg[e^{\f{\pi \tom}{a}}\,e^{- i \tom \D \t}
+\,e^{-\f{\pi \tom}{a}}\,e^{i \tom \D \t}\bigg]~.
\label{wight_2dl_1_proper}
\eeq

\subsection{Response function in $(1+1)$ dimensions}
Now we turn our focus to obtaining the response function per unit time of the detector in $(1+1)$ dimensions where we use Eq. (\ref{wight_2dr_1_proper}) in the Eq. (\ref{power_1})   
which leads us to the response function of the detector as follows:
%%%%%
\beq
I^{R}(\D E)=\,\f{1}{4}\int_{-\infty}^{\infty} dk\,\f{\bigg(e^{\f{\pi \tom}{a}}\d(\D E + \tom)+e^{-\f{\pi \tom}{a}}\d(\D E - \tom)\bigg)}{\tom \,{\rm Sinh}\left(\f{\pi \tom}{a}\right)\,f^2(l^2 \tom^2,\,l^2 k^2)}~.
\eeq
Now we know that $\tom=\,|k|$, thus $\tom=k$ for $k >0$ and $\tom=\,-k$ for $k < 0$. The first term will be zero as $\D E$ and $\tom$ both are positive quantities. 
Therefore the above integral over $k$ can be written as : 
%%%%%
\beq
I^{R}(\D E)=\,\f{1}{2}\int_{0}^{\infty}\,\f{d \tom\,\,e^{-\f{\pi \tom}{a}}\,\,\d(\D E - \tom)}{\tom \,{\rm Sinh}\left(\f{\pi \tom}{a}\right)\,f^2(l^2 \tom^2,\,l^2 k^2)}=\,\f{1}{\D E\,f^2(l^2 \D E^2)\bigg(e^{2\pi \D E/a}-\,1\bigg)}~.
\eeq
We expand the $f^2(l^2 \D E^2)$ as follows, 
\beq
\f{1}{f^2(l^2 \D E^{2})}=\,\sum_{n=0}^{\infty}\,\a_n\,\left( l^2\,\D E^2\right)^n
\label{series_f_1}
\eeq
%%%%%
where the coefficient $\a_0$ is fixed by $\a_0=1$ to recover the local case. Using the series expansion of the function $f$, we get the response function
%%%%
\beq
I^{R}(\D E)=\,\sum_{n=0}^{\infty}\,\f{\a_n\,l^{2 n}\,\D E^{2 n - 1}}{\bigg(e^{2\pi \D E/a}-\,1\bigg)}~.
\label{BRM222}
\eeq
%%%%

%%%%%%%%%%%%%%%%%%%%%%%
\section{FINDING THE NONLOCAL WIGHTMAN FUNCTION IN $(1+3)$ DIMENSIONS USING UNRUH QUANTIZATION}\label{wight_3d_nonlocal}\label{nl_wight_3d}
%%%%
From Eq. (\ref{scalar_nl_3d_r}), we write
\beq
\phi_{\rm NL}^{R}=\,\int_{0}^{\infty} d\tom\,\int_{-\infty}^{\infty} \f{d^2k_{\bot}}{\sqrt{2\,{\rm Sinh}(\f{\pi\tom}{\a})}}\bigg[p^{1}_{k}\,e^{\f{\pi\,\tom}{2\,a}}\,^{R}u_{\tom k_{\bot}}(\eta,\xi,x_{\bot})+\,p^{2}_{k}\,e^{-\f{\pi\,\tom}{2\,a}}\,^{R}u^{*}_{\tom -k_{\bot}}(\eta,\xi,x_{\bot})\bigg]+\,{\rm H.c.}~.
\label{wight_3d_1}
\eeq
%%%%
Wightman function in $(1+3)$ dimensional RRW can be written as 
\beq
{}^{R}G_{W}(x_{1},\,x_{2 })=\,\bra{0_M}{\phi}^R_{\rm NL}(x_{1})\,{\phi}^{R}_{\rm NL}(x_{2 })\ket{0_M}~.
\eeq
Using Eq. (\ref{wight_3d_1}) we obtain
\beq
{}^{R}G_{W}(x_{1},\,x_{2 })&=&\,\int_{0}^{\infty} d\tom_1\,d\tom_2\,\int \f{d^2k_{\bot 1}\,d^2k_{\bot 2}}{\sqrt{4\,{\rm Sinh}(\f{\pi \tom_1}{a})\,{\rm Sinh}(\f{\pi \tom_2}{a})}}\bigg[\bra{0_M}p^{1}_{\tom_1 k_{\bot 1}}\,p^{1^{\dagger}}_{\tom_2 k_{\bot 2}}\ket{0_M}\,e^{\f{\pi\,\tom_1}{2\,a}}\,e^{\f{\pi\,\tom_2}{2\,a}}\,^{R}u_{\tom_1 k_{\bot 1}}\,^{R}u^{*}_{\tom_2 k_{\bot 2}}\nonumber\\
&+&\bra{0_M}p^{2}_{\tom_1 k_{\bot 1}}\,p^{2^{\dagger}}_{\tom_2 k_{\bot 2}}\ket{0_M}\,e^{-\f{\pi\,\tom_1}{2\,a}}\,e^{-\f{\pi\,\tom_2}{2\,a}}\,^{R}u^{*}_{\tom_1 -k_{\bot 1}}\,^{R}u_{\tom_2 -k_{\bot 2}}
\bigg]~.
\eeq
Further using Eq. (\ref{commut_3d_1}) one gets
\beq
{}^{R}G_{W}(x_{1},\,x_{2 })=\,\frac{1}{2}\int_{0}^{\infty} d\tom_1\,d\tom_2\,\int \f{d^2k_{\bot 1}\,d^2k_{\bot 2}}{\sqrt{{\rm Sinh}(\f{\pi \tom_1}{a})\,{\rm Sinh}(\f{\pi \tom_2}{a})}}\,\f{\d(\tom_1-\tom_2)\,\d^2(k_{\bot 1}-k_{\bot 2})}{f(k_1)\,f(k_2)}\nonumber\\
\bigg[e^{\f{\pi\,\tom_1}{2\,a}}\,e^{\f{\pi\,\tom_2}{2\,a}}\,^{R}u_{\tom_1 k_{\bot 1}}\,^{R}u^{*}_{\tom_2 k_{\bot 2}}+e^{-\f{\pi\,\tom_1}{2\,a}}\,e^{-\f{\pi\,\tom_2}{2\,a}}\,^{R}u^{*}_{\tom_1 -k_{\bot 1}}\,^{R}u_{\tom_2 -k_{\bot 2}}\bigg]~.
\eeq
%%%
Here we write $f(k_{1, 2})=f\bigg[\f{l^2\tom^{2}_{1,2}}{a^2},\,l^2(\tom^{2}_{1,2}-e^{2 a \xi_{1,2}}\,k_{\bot 1, 2}^{2}),\,l^2 k_{\bot 1, 2}^{2}\bigg]$. Performing the integrations on $\tom_2$ and $k_{\bot 2}$ and then using Eq. (\ref{rindler_3d_1}) we get
\beq
{}^{R}G_{W}(x_{1},\,x_{2 })=\,\f{1}{8 \pi^4 a}\int_{0}^{\infty} d\tom\,\int \f{d^2k_{\bot}\,K_{\f{i\tom}{a}}\bigg(\f{|k_\bot |\,e^{a\xi_1}}{a}\bigg)\,K_{\f{i\tom}{a}}\bigg(\f{|k_\bot |\,e^{a\xi_2}}{a}\bigg)\,e^{i k_{\bot}(x_{\bot 1}-x_{\bot 2})}}{f\bigg[\f{l^2\tom^{2}}{a^2},\,l^2(\tom^{2}-e^{2 a \xi_1}\,k_{\bot}^{2}),\,l^2 k_{\bot}^{2}\bigg]\,f\bigg[\f{l^2\tom^{2}}{a^2},\,l^2(\tom^{2}-e^{2 a \xi_2}\,k_{\bot}^{2}),\,l^2 k_{\bot}^{2}\bigg]\,}\nonumber\\
\times \bigg[e^{\f{\pi\,\tom}{a}}\,e^{-i\tom(\eta_1-\eta_2)}+\,e^{-\f{\pi\,\tom}{a}}\,e^{i\tom(\eta_1-\eta_2)}\bigg]~.
\eeq
%%%
We take $x_{\bot 1}=\,x_{\bot 2}=\,c_1$ where $c_1$ is a constant and impose the proper frame condition 
as mentioned in Appendix \ref{nl_wight_2d}. This parametrization leads us to the required form of the Wightman function in $(1+3)$ dimensional RRW as depicted in Eq. (\ref{wight_3d_2}). 
%%%%
%%%%%
Furthermore following the discussion in Appendix \ref{nl_wight_2d} for the LRW, one obtains  the positive frequency Wightman function in the LRW as written in Eq. (\ref{wight_3d_2}).  
\section{FINDING THE NONLOCAL WIGHTMAN FUNCTION IN $(1+3)$ DIMENSIONS USING RINDLER QUANTIZATION}\label{wight_3d_nonlocal}
%\label{nl_wight_rindler_3d}
%
%
We write the scalar field solution in RRW by using the Rindler quantization as below [see Eq.(\ref{field_rind_nl_1})]: 
\beq
\phi_{\rm NL}^{R} (x)&=&\,\int_{0}^{\infty}d\tom\int d^{2}k_{\bot}\, \big[{}^{R}c_{\tom k_{\bot}}\,{}^{R}u_{\tom k_{\bot}}(\eta,\xi,x_{\bot})+{}^{R}c_{\tom k_{\bot}}^{^{\dagger}}\,^{R}u^{*}_{\tom k_{\bot}}(\eta,\xi,x_{\bot})\big]~.
\eeq
Therefore using the Rindler quantization the nonlocal Wightman function becomes
\beq
{}^{R}G_{W}(x_{1},\,x_{2 })=\int_{0}^{\infty} d\tom_1\,d\tom_2\int \f{d^2k_{\bot 1} d^2k_{\bot 2}}{f(k_1)\,f(k_2)}&\bra{0_M}\bigg[{}^{R}b_{\tom_1 k_{\bot 1}}{}^{R}b_{\tom_2 k_{\bot 2}}\,{}^{R}u_{\tom_{1} k_{\bot 1}}(\eta_1,\xi_1,x_{\bot 1})\,{}^{R}u_{\tom_{2} k_{\bot 2}}(\eta_2,\xi_2,x_{\bot 2})\nonumber\\
&+ {}^{R}b_{\tom_1 k_{\bot 1}}\,{}^{R}b^{\dagger}_{\tom_2 k_{\bot 2}}\,{}^{R}u_{\tom_{1} k_{\bot 1}}(\eta_1,\xi_1,x_{\bot 1})\,{}^{R}u_{\tom_{2} k_{\bot 2}}^{*}(\eta_2,\xi_2,x_{\bot 2}) \nonumber\\
&+{}^{R}b^{\dagger}_{\tom_1 k_{\bot 1}}\,{}^{R}b_{\tom_2 k_{\bot 2}}\,{}^{R}u_{\tom_{1} k_{\bot 1}}^{*}(\eta_1,\xi_1,x_{\bot 1})\,{}^{R}u_{\tom_{2} k_{\bot 2}}(\eta_2,\xi_2,x_{\bot 2})\nonumber\\
&+{}^{R}b^{\dagger}_{\tom_1 k_{\bot 1}}{}^{R}b^{\dagger}_{\tom_2 k_{\bot 2}}\,{}^{R}u_{\tom_{1} k_{\bot 1}}^{*}(\eta_1,\xi_1,x_{\bot 1})\,{}^{R}u_{\tom_{2} k_{\bot 2}}^{*}(\eta_2,\xi_2,x_{\bot 2})\bigg]\ket{0_M}~.\nonumber\\
\label{wight_rindler_1}
\eeq
We write the Rindler operators in terms of the Minkowski operators by using the Bogoliubov transformation as follows, 
\beq
{}^{R}b_{\tom k_{\bot}}=\int_{-\infty}^{\infty}\,d^3\vec{k'} \,\,\big[{}^{R}\alpha_{\vec{k} \vec{k'}}\, a_{\vec{k'}} +{}^{R}\b_{\vec{k} \vec{k'}}\, a^{\dagger}_{\vec{k'}}\big]~.
\label{bogolibov_1}
\eeq
Here,
\beq
{}^{R}\alpha_{\vec{k} \vec{k'}}={}^{R}\alpha_{\tom k_{z}^{'}k_{\bot}^{'}}&=&\f{e^{\pi \tom/2 a}}{\sqrt{4 \pi k_{0}^{'}a \sinh(\pi \tom/a)}}\,\Bigg(\f{k_{0}^{'}+k_{z}^{'}}{k_{0}^{'}-k_{z}^{'}}\Bigg)^{-i \tom/a}\,\d^2 (k_\bot -k_{\bot}^{'})\nonumber\\
{}^{R}\b_{\vec{k} \vec{k'}}={}^{R}\b_{\tom k_{z}^{'}k_{\bot}^{'}}&=&\f{- \,e^{\pi \tom/2 a}}{\sqrt{4 \pi k_{0}^{'}a \sinh(\pi \tom/a)}}\,\Bigg(\f{k_{0}^{'}+k_{z}^{'}}{k_{0}^{'}-k_{z}^{'}}\Bigg)^{-i \tom/a}\,\d^2 (k_\bot + k_{\bot}^{'})~,
\eeq
are known as the Bogoliubov coefficients.  It can be shown that the sum of the first two terms in Eq. (\ref{wight_rindler_1}) yield zero. 
Using Eq. (\ref{bogolibov_1}), we examine the third term of (\ref{wight_rindler_1}) as 
\beq
\bra{0_M}{\rm 3rd\, term}\ket{0_M} &=&\bigintsss_{0}^{\infty} d\tom_1 d\tom_2 \bigintsss d^2k_{\bot 1}\,d^2k_{\bot 2}\bigintsss  d^3k' d^3k''\,
\f{\bra{0_M}a_{\vec{k'}}a_{\vec{k''}}^{\dagger}\ket{0_M}\,{}^{R}\b^{*}_{\vec{k} \vec{k'}}\,{}^{R}\b_{\vec{k} \vec{k'}}\,{}^{R}u_{\tom_{1} k_{\bot 1}}^{*}{}^{R}u_{\tom_{2} k_{\bot 2}}}{f(k_1)\,\,f(k_2)}\nonumber\\
&=&\bigintsss_{0}^{\infty} d\tom_1\,d\tom_2 \bigintsss d^2k_{\bot 1}\,d^2k_{\bot 2}\,d^3k'\,d^3k''\,\f{\o_{\vec{k'}}\,\d^3 (\vec{k'}-\vec{k''})\,{}^{R}\b^{*}_{\vec{k} \vec{k'}}\,{}^{R}\b_{\vec{k} \vec{k'}}\,{}^{R}u_{\tom_{1} k_{\bot 1}}^{*}{}^{R}u_{\tom_{2} k_{\bot 2}}}{\sqrt{\o_{\vec{k'}}\,\o_{\vec{k''}}}\,\,f(k_1)\,f(k_2)}~.\nonumber\\
\label{3rdterm_1}
\eeq
We perform the integral over $k''$, $k_{\bot 2}$ and $k_{\bot}^{'}$. Due to the delta functions in the integrals, we fix $\vec{k''}=\vec{k'}$, $k_{\bot 2}= -\,k_{\bot}^{'}$ and $k_{\bot}^{'}=k_{\bot 1}$, which yields
\beq
\bra{0_M}{\rm 3rd\, term}\ket{0_M}=
\f{1}{8 \pi^4 a} \bigintsss_{0}^{\infty} d\tom_1\,d\tom_2 \bigintsss d^2k_{\bot 1} \bigintsss \f{d k_{z}^{'}}{2 \pi k_{0}^{'} a}\Bigg(\f{k_{0}^{'}+k_{z}^{'}}{k_{0}^{'}-k_{z}^{'}}\Bigg)^{\f{i}{a}(\tom_1 - \tom_2)}\f{e^{- \f{\pi}{2 a}(\tom_1 + \tom_2)} e^{i (\tom_1 \t_1 - i \tom_2 \t_2)}}{f(k_1)\,\,f(k_2)}\nonumber\\
K_{\f{- i\tom_1}{a}}\bigg(\f{|k_{\bot 1}|}{a}\bigg) K_{\f{i\tom_2}{a}}\bigg(\f{|k_{\bot 1}|}{a}\bigg)
\label{3rdterm_2}~.
\eeq
In the above equation we consider the proper frame of the Rindler observer and replace $\xi_{1,2}=x_{\bot 1,2}=0$ and $\eta_{1,2}=\t_{1,2}$. Following the Sec. E of \cite{Crispino} we take  $\theta (k_{z}^{'})={\rm log}\Bigg(\f{k_{0}^{'}+k_{z}^{'}}{k_{0}^{'}-k_{z}^{'}}\Bigg)$ and use the result of the integral, 
$\bigintss_{-\infty}^{\infty} \f{dk_{z}^{'}}{2 \pi k_{0}^{'} a} e^{i \theta (k_{z}^{'})(\tom_1 - \tom_2)/a}=\d(\tom_1 - -\tom_2)$ in Eq. (\ref{3rdterm_2}). 
Subsequently the integration over $\tom_2$ yields
\beq
\bra{0_M}{\rm 3rd\, term}\ket{0_M} =
\f{1}{8 \pi^4 a} \bigintsss_{0}^{\infty} d\tom \bigintsss d^2k_{\bot} \,e^{-\pi \tom/a}\Bigg{|}K_{\f{i\tom}{a}}\bigg(\f{|k_{\bot}|}{a}\bigg)
\Bigg{|}^2\,\f{e^{i \,\tom \D \t}}{[f(k)]^2}
\label{3rd_f2}~,
\eeq
where we replace $k_{\bot 1}=k_{\bot}$. Following the above analysis the fourth term of Eq. (\ref{wight_rindler_1}) yields
\beq
\bra{0_M}{\rm 4th\, term}\ket{0_M}= \f{1}{8 \pi^4 a} \bigintsss_{0}^{\infty} d\tom \bigintsss d^2k_{\bot} \,e^{\pi \tom/a}\Bigg{|}K_{\f{i\tom}{a}}\bigg(\f{|k_{\bot}|}{a}\bigg)
\Bigg{|}^2\,\f{e^{-i \,\tom \D \t}}{[f(k)]^2}~.
\label{4th_f2}
\eeq
Summing up the Eqs. (\ref{3rd_f2}) and (\ref{4th_f2}) we obtain
\beq
{}^{R}G_{W}(x_{1},\,x_{2 })&=&\int_{0}^{\infty} d\tom_1 \int \f{d^2k_{\bot}}{[f(k)]^2}\, \Bigg{|}K_{\f{i\tom}{a}}\bigg(\f{|k_{\bot}|}{a}\bigg)
\Bigg{|}^2\,\Big[e^{\pi \tom/a} \,e^{-i \,\tom \D \t}+ e^{-\pi \tom/a}\, e^{i \,\tom \D \t} \Big]~.
\label{wight_final}
\eeq
Note that Eq. (\ref{wight_final}) and (\ref{wight_3d_2}) match each other. Thus Rindler quantization also produces the time translationally invariant nonlocal Wightman function in the proper frame of the accelerated observer. In this approach one needs to know the explicit forms of the Bogoliubov coefficients, whereas these are not explicitly required through the Unruh quantization method.

\end{document}